\documentclass[11pt]{article}
\usepackage[final]{acl}
\usepackage{times}
\usepackage{latexsym}
\usepackage[T1]{fontenc}
\usepackage[utf8]{inputenc}
\usepackage{microtype}
\usepackage{inconsolata}
\usepackage{graphicx}

\usepackage{balance}
\usepackage{amsmath}
\usepackage{array}
\usepackage{verbatim}
\usepackage{booktabs}
\usepackage{enumitem}
\usepackage{framed}
\usepackage{bm}
\usepackage{xcolor}
\usepackage{multirow}

\title{AutoUE: Automated Generation of 3D Games in Unreal Engine via Multi-Agent Systems}

\author{
 \textbf{Lei Yin\textsuperscript{1, \thanks{Equal contribution.}}},
 \textbf{Wentao Cheng\textsuperscript{1, \footnotemark[1]}},
 \textbf{Zhida Qin\textsuperscript{1, \thanks{Corresponding author.}}},
 \textbf{Tianyu Huang\textsuperscript{1}},
 \textbf{Yidong Li\textsuperscript{2}},
 \textbf{Gangyi Ding\textsuperscript{1}}
\\
\\
 \textsuperscript{1}Beijing Institute of Technology,
 \textsuperscript{2}Beijing Jiaotong University
\\
 {\tt \{3120241020, zanderqin, huangtianyu, dgy\}@bit.edu.cn,}
\\
 {\tt wentao.cheng23@gmail.com, ydli@bjtu.edu.cn}
}

\begin{document}
\maketitle
\begin{abstract}
Automatically generating 3D games in commercial game engines remains a non-trivial challenge, as it involves complex engine-related workflows for generating assets such as scenes, blueprints, and code. To address this challenge, we propose a novel multi-agent system, \textbf{AutoUE}, which coordinates multiple agents to end-to-end generate 3D games, covering model retrieval, scene generation, gameplay and interaction code synthesis, and automated game testing for evaluation. In order to mitigate tool-use hallucinations in LLMs, we introduce a retrieval-augmented generation mechanism that grounds agents with relevant UE tool documentation. Additionally, we incorporate game design patterns and engine constraints into the code generation process to ensure the generation of correct and robust code. Furthermore, we design an automated play-testing pipeline that generates and executes runtime test commands, enabling systematic evaluation of dynamic behaviors. Finally, we construct a game generation dataset and conduct a series of experiments that demonstrate AutoUE's ability to generate 3D games end-to-end, and validate the effectiveness of these designs.\footnote{The code is released at \url{https://github.com/Pluto156/AutoUE}.}
\end{abstract}

\section{Introduction}
Digital games, as one of the most influential forms of interactive media, play an essential role in entertainment and simulation \cite{martins2025llm, park2023generative}. Due to the need to design complex visual environments and system logic, the game creation remains a highly creative yet labor-intensive process.

Recent advances in Large Language Models (LLMs) \cite{guo2025deepseek} and generative models \cite{ho2020denoising} have motivated studies on automated game generation from natural language descriptions \cite{chegamegen, valevskidiffusion, dai2024procedural}, aiming to reduce development barriers. A promising direction is to integrate LLMs with game engines to generate in-engine assets \cite{maleki2024procedural, buongiorno2024pangea}, which offers better controllability, editability, and practical value for game development.

\begin{figure}[t]
    \vspace{-0.1in}
    \centering
    \includegraphics[width=7.6cm]{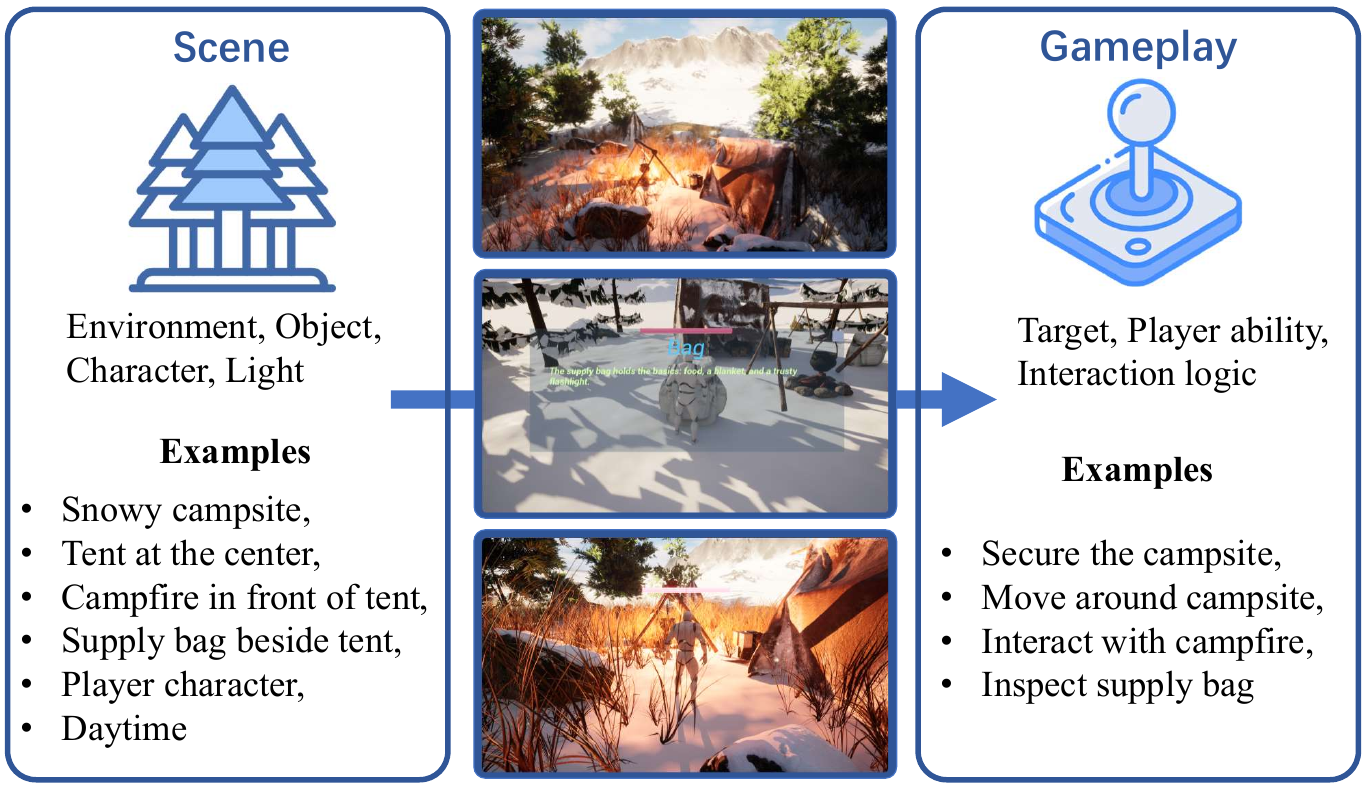}
    \captionsetup{skip=1pt}
    \caption{An example of game scene and gameplay.}
    \label{fig:Intro}
    \vspace{-0.2in}
\end{figure}

Despite these advances, automatically generating 3D games in game engines remains a non-trivial challenge ~\cite{earle2025dreamgarden}. Compared with 2D or simple game settings \cite{todd2024gavel, hu2024game, sudhakaran2023mariogpt}, 3D games require more complex spatial layouts, gameplay behaviors, and coordination across multiple subsystems. Recent works have integrated LLMs with commercial engines such as Unreal Engine (UE) and Unity to generate 3D scenes ~\cite{songtang2025unrealllm}, gameplay logic ~\cite{wayama2022verifying}, and engine code~\cite{yang202590}. However, most of these efforts focus on isolated individual components, making it difficult to generate a complete, playable 3D game end-to-end. 

To address these challenges, we propose \textbf{AutoUE}, a novel multi-agent system that coordinates agents for model retrieval, scene generation, gameplay code, interactive object, and automated play-testing to produce complete 3D games in UE.

Specifically, the model retrieval agent builds an embedding database over 858K 3D models, enabling efficient semantic retrieval of relevant model assets. Based on the retrieved models, the scene generation agent uses UE's built-in tool Procedural Content Generation (PCG) to generate a visually editable layout graph that produces a coherent 3D scene. Then, the gameplay code agent follows common game-development design patterns and respects engine-specific constraints to generate modular gameplay C++ code. Building on these modules, the interactive object agent derives object interaction flows and implements the corresponding C++ code, enabling in-scene entities to provide player interactions consistent with the description. Finally, the automated play-testing agent automatically generates and executes runtime test commands to evaluate the scene and gameplay behaviors.
These agents communicate via structured specifications to convey contextual information and ensure consistency across the system.

Our system has the following innovations:

(1) Compared to other engines, UE provides comprehensive built-in tools for efficient game development. To mitigate tool-use hallucinations by LLMs, we propose a retrieval-augmented generation (RAG) based mechanism that retrieves relevant information from tool documents and provides the agent with accurate usage instructions.

(2) Existing methods often overlook design patterns in game development, resulting in unconstrained code that is difficult to maintain and extend. To address this limitation, we generate functional gameplay module code and design a module management framework based on common practices. This enables the system to produce reusable and composable gameplay code, which can be invoked by interactive objects in a controlled manner.

(3) Prior work evaluates generated assets via human inspection or static checks. However, a complete game is defined by dynamic gameplay and interaction behaviors, making objective evaluation without human effort challenging. We introduce an automated play-testing pipeline that generates test commands and implements a Model Context Protocol (MCP) to execute them at runtime, enabling automated and systematic evaluation.

Finally, we construct a dataset of 20 game-generation tasks and demonstrate that our method produces higher-quality 3D scenes than existing approaches. Meanwhile, we design multiple metrics that evaluate dynamic gameplay behaviors based on the generated gameplay and interaction code as well as runtime logs, providing a benchmark for end-to-end 3D game generation in game engines.

Overall, our contributions are summarized as:
\begin{itemize} [leftmargin=*, itemsep=0pt, parsep=0pt, topsep=2pt, partopsep=0pt]
\item We present a practical end-to-end multi-agent system for generating 3D games in UE. The system collaboratively constructs a coherent 3D scene, synthesizes robust gameplay and interaction code, and performs automated play-testing.
\item We design a RAG mechanism that provides agents with tool-usage guidance, thereby reducing hallucinations for LLMs. During code generation, we incorporate game design patterns and engine constraints to improve the maintainability and extensibility of the generated code. 
\item We construct a game-generation dataset and establish an automated play-testing pipeline, providing a benchmark for 3D game generation. It offers a solid foundation for multi-agent game generation, and the overall system provides an extensible infrastructure for future work.
\end{itemize}

\section{Related Work}
Existing research on applying agents to games can be divided into two types:
\subsection{LLMs as Game Playing Agents}
One line of work treats LLMs as game playing agents and uses games as testbeds to improve the agents's ability to process complex observations.
Early studies primarily focus on text-based games: \citet{hausknecht2020interactive} proposes an evaluation framework for interactive fiction and adopts a template-based action space to reduce complexity. 
% CALM \cite{yao2020keep} generates a compact set of candidate actions and re-ranks them with reinforcement learning to mitigate the combinatorial explosion of actions. 
% ReAct \cite{yao2022react} interleaves thought–action–observation sequences, allowing external feedback to guide subsequent planning and decisions. 
Recent studies have further applied LLMs to open-world 3D games. 
DEPS \cite{wang2023describe} conditions planning on environment feedback and sub-goal selection to robustly complete diverse Minecraft tasks. 
Voyager \cite{wangvoyager} combines an expanding skill library with a feedback-driven iterative loop to achieve open-ended exploration in Minecraft. In addition, recent work has also explored agents in a wide range of game genres, including competitive \cite{wang2025explore} and action \cite{chen2024can} games. The latest advances \cite{tan2025lumine} train foundation models on large-scale gameplay data, enabling generalization to unseen games. 

\begin{figure*}[t]
    \vspace{-0.1in}
    \centering
    \includegraphics[width=15.6cm]{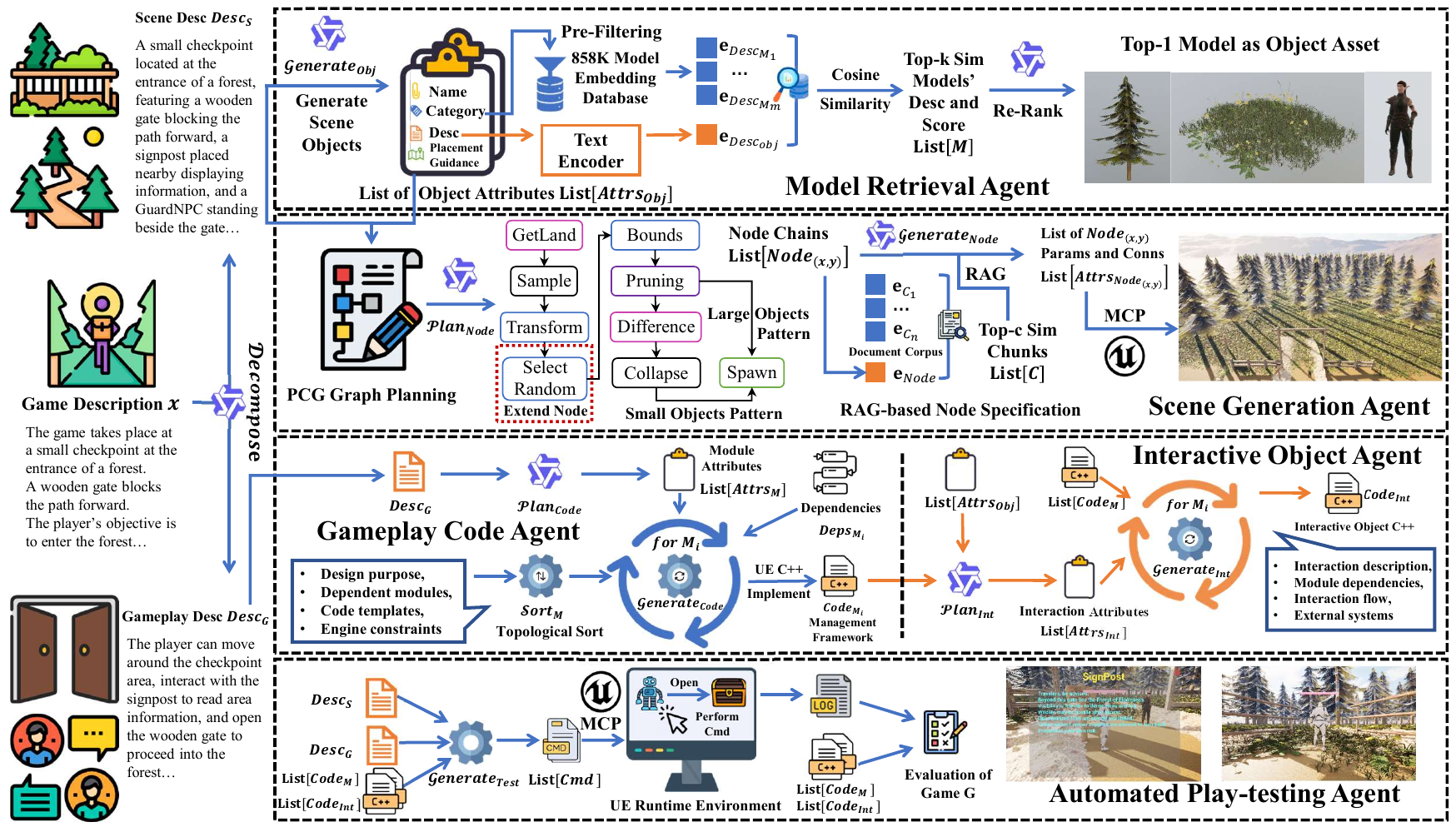}
    \caption{The overall framework of our proposed multi-agent system AutoUE. Blue arrows indicate the workflow.}
    \label{fig:AutoUE}
    \vspace{-0.18in}
\end{figure*}

\subsection{LLMs as Game Generation Agents}
Another line of work uses LLMs to generate game content, including text/video-based and engine-integrated game generation. The former generates game experiences in the form of text narratives \cite{cote2018textworld}, scripted interactions \cite{sun2023language}, or video-like content \cite{bruce2024genie}. Such games are easier to synthesize end-to-end, but they lack free-form spatial interactions and developer-editable content.
Therefore, recent studies couple LLMs with game engines to generate in-engine assets for modern game development. Sketch2Scene \cite{xu2024sketch2scene} uses diffusion models to turn user sketches into isometric 2D scene guides, and feeds guides into procedural generation to build 3D game scenes. 3Dify \cite{hayashi20253dify} uses RAG and MCP to invoke tools like Blender, Unity, and UE to create 3D scenes with iterative user feedback. UnrealLLM \cite{songtang2025unrealllm} achieves automated high-quality 3D scene generation in UE by leveraging multi-modal asset retrieval and PCG powered by expert knowledge. \citet{hassan2025automated} parse game design documents and fine-tune LLMs to generate modular, compilable, and design-consistent Unity C\# game template prototypes. UniGen \cite{yang202590} builds a multi-agent framework to generate Unity C\# code, attach the code to components, and support iterative debugging. These studies focus on a single stage of the pipeline in isolation. Due to the tightly coupled nature of game engine workflows, these components are often not trivially composable. DreamGarden~\cite{earle2025dreamgarden} takes a step toward end-to-end generation by using hierarchical planning and iteratively construct a game prototype in UE. Nevertheless, its generated results remain limited in overall quality and lack robustness when facing complex requirements, highlighting the need for a more reliable, workflow-aligned system for complete 3D game generation in game engines.

\section{Methodology}
\subsection{System Overview}
Figure~\ref{fig:AutoUE} illustrates our multi-agent system, which consists of five LLMs agents: (1) Model Retrieval Agent, (2) Scene Generation Agent, (3) Gameplay Code Agent, (4) Interactive Object Agent, and (5) Automated Play-testing Agent.

We adopt the model context protocol as a standardized interface between our agents and UE. When an agent produces structured specifications that need to be integrated into UE, it invokes the engine tools or code exposed via our MCP implementation to import the outputs into the project and execute the corresponding application workflow. As MCP primarily serves as an engineering integration layer, we do not further elaborate on its implementation details in the following sections.

\subsection{Problem Formulation} \label{sec:Problem Formulation}
Given a game description $x$, our goal is to construct a complete UE game $G$ that consists of a coherent 3D scene, executable gameplay code, interactive object logic, and automated testing instructions.

Since $x$ often interleaves scene content with gameplay intent, it blurs the line between what should be built and what should happen, potentially confusing downstream agents. Therefore, we first decompose $x$ into a scene description $Desc_S$ and a gameplay description $Desc_G$:
\begin{equation}
    Desc_S, Desc_G = \mathcal{D}ecompose(x),
\end{equation}
where $Desc_S$ defines the scene context and key objects, while $Desc_G$ defines the gameplay mechanics and player interactions; and $\mathcal{D}ecompose$ denotes the prompt used to query LLMs. The prompt details are provided in Appendix Figure~\ref{fig:prompt:decompose}.

\subsection{Model Retrieval Agent} \label{sec:Model Retrieval Agent}
As the fundamental entities in the game world, scene objects are not only crucial for the visual presentation, but also provide the essential basis for interaction modeling. Based on $Desc_S$, we construct a list of key scene object along with their associated attributes, represented as:
\begin{equation}
    \text{List}[Attrs_{Obj}] = {\mathcal{G}enerate}_{Obj}(Desc_S),
\end{equation}
where the object attributes $Attrs_{Obj}$ contain name, category, description, and spatial placement guidance; and the prompt details of ${\mathcal{G}enerate}_{Obj}$ are provided in Appendix Figure~\ref{fig:prompt:generate_obj}.

The $Attrs_{Obj}$ are then used in our retrieval pipeline, which is built on TexVerse ~\cite{zhang2025texverse}, a repository with over 858K text-described 3D models from Sketchfab. To efficiently retrieve relevant models, we use a text encoder to create an embedding database from the model descriptions. 

Due to the large number of models, we first apply a pre-filtering process using Sketchfab's 18 categories to narrow down the search according to the category in $Attrs_{Obj}$. Next, we calculate the cosine similarity between the encoded object description in $Attrs_{Obj}$ and the embedding of filtered models to retrieve the top-k models, formulated as:
\begin{equation}
    \resizebox{.88\hsize}{!}{$\text{List}[M] = \mathcal{S}ort\left(\left[\mathcal{S}im(\mathbf{e}_{Desc_{Obj}}, \mathbf{e}_{Desc_{M_i}}) \right]_{i=1}^{m} \right)[:k],$}
\end{equation}
where $\text{List}[M]$ is the list of model description and similarity score of the top-$k$ retrieved models, $\mathcal{S}ort$ denotes the function that sorts values in descending order, $\mathcal{S}im$ denotes the cosine similarity function, $\mathbf{e}_{Desc_{Obj}}$ and $\mathbf{e}_{Desc_{M_i}}$ are the embeddings of the object description and the $i$-th model description, and $m$ is the number of models after pre-filtering.

Then, we use LLMs to re-rank $\text{List}[M]$ based on $Attrs_{Obj}$. The top-1 model after reranking is used as the retrieved 3D model for the object.

\subsection{Scene Generation Agent} \label{sec:Scene Generation Agent}
After retrieving high-quality 3D object models, we stably build a coherent UE scene based on these assets. Instead of generating hard-coded coordinates, we leverage PCG to synthesize the scene layout.
PCG represents scene construction as a visually configurable graph, connecting modular nodes that support functions such as surface sampling, point transformation, and bounds modification. These nodes make PCG a natural fit for $Desc_S$, as they fulfill most of the abstract placement intentions and can be extended to meet additional needs. 

\subsubsection{PCG graph planning} \label{sec:PCG graph planning}
Based on $Desc_S$ and $\text{List}[Attrs_{Obj}]$, we generate one PCG node chain per scene object. Each chain is described by an ordered list of node types with explicit editor coordinates, denoted as:
\begin{equation}
    \resizebox{.88\hsize}{!}{$\text{List}[Node_{(x,y)}] \!=\! {\mathcal{P}lan}_{Node}(Desc_S, Attrs_{Obj}),$}
\end{equation}
where $\text{List}[Node{(x,y)}]$ denotes a PCG node chain, whose each element is a node of type $Node$ located at editor coordinates $(x,y)$, and the prompt details of ${\mathcal{P}lan}_{Node}$ are provided in Appendix Figure~\ref{fig:prompt:plan_node}.

In practice, unconstrained PCG graph generation leads to unstable or hard-to-debug procedural behaviors. Therefore, we classify the object placement into two canonical PCG patterns, each capturing a frequent production need:
\begin{itemize} [leftmargin=*, itemsep=0pt, parsep=0pt, topsep=2pt, partopsep=0pt]
\item Large objects (e.g., buildings) require stable sampling and spacing. The chain ends with a direct spawn step after pruning and bounds control to avoid overlaps and out-of-bound placements.
\item Small scatter objects (e.g., trees, rocks) must avoid intersecting major actors. We add exclusion nodes to subtract forbidden regions, improving collision avoidance and visual coherence.
\end{itemize}
This design enables production-level stability, provides flexibility through parameters (e.g., density, bounds), and can meet a broader range of requirements through extended modes.

\subsubsection{RAG-based node specification} \label{sec:RAG-based node specification}
The planned list only specifies which nodes appear, but it lacks the parameters and pin connections. To avoid hallucinating these details, we adopt RAG over the full document corpus: we segment all available documents into a unified set of text chunks and retrieve the most relevant chunks for each target node. In practice, source material may include cross-references across documents and sections. A strictly indexing scheme is expensive to maintain, scales poorly with new nodes, and may lead to excessive context during generation. Therefore, we use semantic retrieval to fetch the minimal guidance needed for each node, represented as:
\begin{equation}
    \resizebox{.88\hsize}{!}{$\text{List}(C) = \mathcal{S}ort\left(\left[\mathcal{S}im(\mathbf{e}_{Node}, \mathbf{e}_{C_i}) \right]_{i=1}^{n} \right)[:c],$}
\end{equation}
\begin{equation}
    \resizebox{.88\hsize}{!}{$Attrs_{Node} \!=\! {\mathcal{G}enerate}_{Node}(Node_{(x,y)}, \text{List}(C)),$}
\end{equation}
where $n$ is the number of chunks, $C_i$ is the $i$-th chunk, $\mathbf{e}_{Node}$ and $\mathbf{e}_{C_i}$ are the embeddings of $Node_{(x,y)}$ and $C_i$, $\text{List}(C)$ is the retrieved top-c chunks, $Attrs_{Node}$ are the parameters and connections for $Node_{(x,y)}$, and the details of ${\mathcal{G}enerate}_{Node}$ are provided in Appendix Fig~\ref{fig:prompt:generate_node}.

The RAG knowledge enables PCG to generate reliable scenes, while also making it easy to introduce new node capabilities by adding document. Finally, we develop MCP tools to convert all $Attrs_{Node}$ into the UE PCG editor. This approach can also be extended to other game development tools, as they typically modular in design and provide the necessary usage information in documents.

\subsection{Gameplay Code Agent} \label{sec:Gameplay Code Agent}
To realize executable gameplay logic, we follow a planned generation pipeline, where the system first analyzes $Desc_G$ to plan the required functional logic modules, and then generates the corresponding UE C++ implementations, formulated as:
\begin{equation}
    \text{List}[Attrs_{M}] = {\mathcal{P}lan}_{Code}(Desc_G),
\end{equation}
{\small\begin{equation}
\begin{array}{c}
    \text{for } M_i \in {\mathcal{S}ort}_{M}(\text{List}[Attrs_{M}]): Code_{M_i} = \\
    {\mathcal{G}enerate}_{Code}(Desc_G, \text{List}[Attrs_{M}], Deps_{M_i}),
\end{array}
\end{equation}}
where $\text{List}[Attrs_{M}]$ is the planned list of module attributes, $M_i$ is the $i$-th module, $\mathcal{S}ort_{M}$ performs a topological sort based on the dependencies between modules, $Code_{M_i}$ is the C++ implementation code of $M_i$, $Deps_{M_i}$ are the generated code of modules that $M_i$ depends on, and the details of $ {\mathcal{P}lan}_{Code}$ and ${\mathcal{G}enerate}_{Code}$ are provided in Appendix Figure~\ref{fig:prompt:plan_code} and Figure~\ref{fig:prompt:generate_code}, respectively.

Specifically, $Attrs_{M}$ contains the design purpose, usage, and dependent modules to provide more context for code generation. We also provide code templates and engine-related constraints (e.g., export/registration macros, include whitelist, logging, and cross-module access) in ${\mathcal{G}enerate}_{Code}$ to improve code generation stability. Additionally, following common game development practices, we design a generic module management framework as MCP for the module code. Together, these designs ensure that gameplay logic can be reliably triggered by in-engine interactions while remaining extensible and consistent with UE.

\subsection{Interactive Object Agent} \label{sec:Interactive Object Agent}
The generated modules (e.g., InventoryModule, DialogueModule) are invoked within the object interaction implementation methods to execute concrete gameplay logic (e.g., pickup, dialogue). To implement the interaction, we first infer which modules should be invoked and the specific order in which their methods should be called, represented as:
\begin{equation}
    \resizebox{.88\hsize}{!}{$\text{List}[Attrs_{Int}] = {\mathcal{P}lan}_{Int}(\text{List}[Attrs_{Obj}], \text{List}[Code_M]),$}
\end{equation}
where $Attrs_{Int}$ denotes the attributes of the interactive object, including interaction description, module dependencies, interaction flow, and whether external systems such as combat or navigation plugins need to be invoked; and the prompt details of ${\mathcal{P}lan}_{Int}$ are provided in Appendix Figure~\ref{fig:prompt:plan_interaction}.

Based on these inferred attributes, we generate the C++ code for each interactive object:
\begin{equation}
    \resizebox{.85\hsize}{!}{$Code_{Int} = {\mathcal{G}enerate}_{Int}(Attrs_{Int}, \text{List}[Code_M]),$}
\end{equation}
where $Code_{Int}$ denotes the generated C++ code for the interactive object, and $\mathcal{G}enerate_{Int}$ ensures consistent interaction logic through the module's context and the interaction flow. The prompt details of $\mathcal{G}enerate_{Int}$ are provided in Appendix Fig~\ref{fig:prompt:generate_interaction}.

\subsection{Automated Play-testing Agent} \label{sec:Automated Play-testing Agent}
After generating the interaction code, we have completed the entire UE game construction. However, there are multiple ways to implement $Desc_S$ and $Desc_G$, and evaluating the generated content becomes a challenge. To address this, we set up an evaluation environment by generating and executing test instructions, denoted as:
\begin{equation}
\begin{aligned}
    & \text{List}[Cmd] = {\mathcal{G}enerate}_{Test}(Desc_S, \\
    & Desc_G, \text{List}[Code_M], \text{List}[Code_{Int}]),
\end{aligned}
\end{equation}
where $\text{List}[Cmd]$ is the testing commands used to evaluate our game $G$, and the prompt details of ${\mathcal{G}enerate}_{Test}$ are provided in Appenidx Figure~\ref{fig:prompt:generate_test}. 

There are two types of testing commands: moving to an object and performing specific interactions. We automatically execute these commands in UE runtime environment through MCP. The runtime screenshots, execution logs, $\text{List}[Node_{(x,y)}]$, $\text{List}[Code_M]$, $\text{List}[Code_{Int}]$, and middle specifications are used for the final evaluation of our generated game $G$, providing a thorough evaluation.

\begin{figure*}[t]
    \vspace{-0.1in}
    \centering
    \includegraphics[width=15cm]{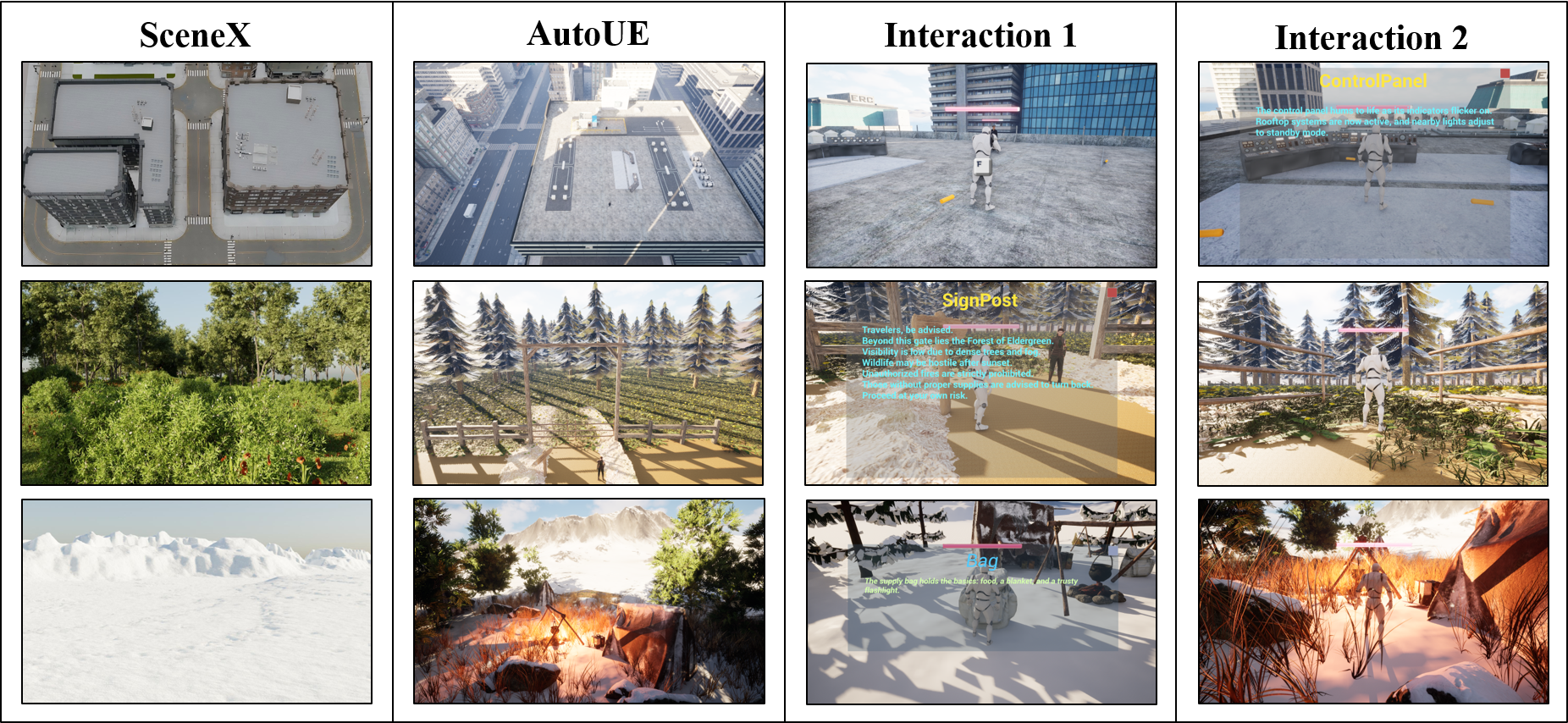}
    \captionsetup{skip=3pt}
    \caption{Visual comparison with SceneX. The first column shows scenes from SceneX; the second column shows the corresponding scenes in AutoUE; the third and fourth columns present AutoUE’s interactions within each scene.}
    \label{fig:Interaction}
    \vspace{-0.15in}
\end{figure*}

\section{Experiments}
We evaluate our multi-agent game generation system \textbf{AutoUE} from five perspectives, corresponding to the key claims made in previous sections:

\begin{itemize}[leftmargin=*, itemsep=2pt, parsep=0pt, topsep=2pt]
\item \textbf{Q1}: Can AutoUE generate complete, playable UE games end-to-end, and what is the overall quality of the generated games?
\item \textbf{Q2}: Compared with existing scene generation baselines, are the scenes generated by AutoUE better in terms of aesthetic quality?
\item \textbf{Q3}: What is the impact of RAG-based node specification on generating valid PCG graphs.
\item \textbf{Q4}: What is the impact of game design patterns on the quality of the generated gameplay code?
\end{itemize}

\subsection{Experimental Setup}
\subsubsection{Dataset}
We construct a benchmark of 20 game generation tasks (denoted as PlayGen-20). These tasks cover common interactions in game design (e.g., pickup, open, dialogue, inspect, and combat). Task details are provided in the Appendix Figure~\ref{fig:dataset:easy}, \ref{fig:dataset:medium}, and \ref{fig:dataset:hard}.

To better evaluate AutoUE under different levels of game design complexity, we further divide PlayGen-20 into three difficulty categories based on scene complexity and interaction complexity:

\begin{itemize}[leftmargin=*, itemsep=2pt, parsep=0pt, topsep=2pt]
\item \textbf{Easy (5 games).} These tasks focus on onboarding and practicing basic interactions. 
The designs include a single or low-complexity interaction, clear feedback with low failure cost. 
This category is well-suited for evaluating end-to-end playability and basic interaction correctness.
\item \textbf{Medium (7 games).} 
These tasks focus on more exploration and collection. The designs include multiple interactive objects, and a slower pace that encourages players to interact more broadly with the environment. 
This category is well-suited for evaluating the ability to organize diverse interactive objects, enrich scene content, and achieve broader interaction coverage.
\item \textbf{Hard (8 games).} These tasks introduce explicit progression constraints and stronger interaction dependencies. 
The designs feature clearer state transitions and tighter coupling between scene objects and gameplay logic. 
This category is well-suited for evaluating robustness under complex interaction flows, including dependency correctness and the reliability of generated code.
\end{itemize}

\subsubsection{Settings}
All experiments are conducted in Unreal Engine 5, which is one of the most advanced and broadly adopted game engines. For LLMs, we use Qwen-Plus as the default backbone for generation and evaluation. In Q2, we instead use GPT-4o only for evaluation to align with baslines. 

\subsection{Overall Game Evaluation (Q1)} \label{sec:Overall Game Evaluation}
\begin{table}[!t]
\vspace{-0.1in}
\centering
\resizebox{0.42\textwidth}{!}{%
\begin{tabular}{ccccc}
\toprule
\textbf{Score} & \textbf{Scene} & \textbf{Gameplay} & \textbf{Visual} & \textbf{Game} \\
\midrule
\textbf{Easy}   & 10.0  & 8.30  & 7.70  & 8.72 \\
\textbf{Medium} & 9.79  & 8.29  & 8.14  & 8.74 \\
\textbf{Hard}   & 9.94  & 8.19  & 7.50  & 8.59 \\
\textbf{All}    & 9.90  & 8.25  & 7.78  & 8.68 \\
\bottomrule
\end{tabular}
}
\captionsetup{skip=3pt}
\caption{The game evaluation results on PlayGen-20.}
\label{tab:expQ1}
\vspace{-0.18in}
\end{table}

To objectively evaluate the overall quality of UE games generated end-to-end by AutoUE, we adopt an LLMs-as-a-judge evaluation method and carefully design the evaluation prompt (the prompt details are provided in Appendix Figure~\ref{fig:eval:overall_game}). 

The LLMs score each game along three dimensions, each on a 1--10 scale: 
(1) \textbf{Scene}: evaluates the correctness of PCG based on build logs and PCG node chains, 
(2) \textbf{Gameplay}: evaluates whether gameplay and interactive logic work as intended, based on gameplay/interaction attributes and code,
(3) \textbf{Visual}: evaluates the visually coherence of game with the description, based on scene and interaction screenshots. The three dimension scores are aggregated into a final \textbf{Game} score using weights of 0.35, 0.35, and 0.3, respectively.

Notably, \textbf{Scene} and \textbf{Gameplay} place strong emphasis on successful build as these two dimensions are used to assess whether AutoUE can generate complete, playable games end-to-end. In contrast, \textbf{Visual} is used to evaluate the quality of games.

As shown shown in Table~\ref{tab:expQ1}, AutoUE achieves strong performance, indicating that the system can stably generate playable UE games end-to-end. 

On the \textbf{Scene} dimension, Hard games score higher than Medium games because Hard tasks involve more complex PCG graphs, and successfully building these graphs yields higher scores. This also indicates that AutoUE can reliably handle more complex scene generation requirements.

On the \textbf{Visual} dimension, Easy games lower than Medium games due to content constrains limit visual richness. Hard games receive the lowest scores primarily because deviations in interactions from the intended behavior degrade visual quality. This demonstrates that the \textbf{Visual} metric is sensitive to the quality of the generated games.

Overall, the \textbf{Game} and \textbf{Gameplay} scores follow the expected difficulty trend, suggesting that the primary challenge at higher difficulty stems from the complexity of gameplay and interaction logic.

\subsection{Scene Generation Evaluation (Q2)}
\begin{table}[!t]
\vspace{-0.1in}
\centering
\resizebox{0.4\textwidth}{!}{%
\begin{tabular}{@{}l c@{}}
\toprule
\textbf{Approach} & \textbf{GAS} \\
\midrule
\multicolumn{2}{@{}l@{}}{\textbf{Generative Baselines}} \\
DreamFusion (ICLR 23) \cite{pooledreamfusion} & 4.83 \\
Magic3D (CVPR 23) \cite{lin2023magic3d} & 6.39 \\
WonderJ (CVPR 24) \cite{yu2024wonderjourney} & 7.38 \\
\midrule
\multicolumn{2}{@{}l@{}}{\textbf{Procedural Baselines}} \\
Infinigen (CVPR 23) \cite{raistrick2023infinite} & 6.61 \\
3D-GPT (3DV 25) \cite{sun20253d} & 6.76 \\
SceneX (AAAI 25) \cite{zhou2025scenex} & 7.31 \\
UnrealLLM (ACL 25) \cite{songtang2025unrealllm} & 7.71 \\
\midrule
\textbf{AutoUE} & \textbf{7.8} \\
\bottomrule
\end{tabular}
}
\captionsetup{skip=3pt}
\caption{The scene generation evaluation results.}
\label{tab:expQ2}
\vspace{-0.18in}
\end{table}

To further evaluate the quality of scenes generated by AutoUE, we follow the evaluation setup of UnrealLLM \cite{songtang2025unrealllm}. Specifically, UnrealLLM uses the GPT Aesthetic Score (GAS) to quantify aesthetic quality by assessing factors such as composition, color harmony, lighting, material and texture fidelity, richness of details, overall coherence, and artistic impact. 
% We compare the GAS of AutoUE with two types of baselines:

% \noindent \textbf{Generative Baselines:} 
% \begin{itemize}[leftmargin=*, itemsep=2pt, parsep=0pt, topsep=2pt]
% \item DreamFusion \cite{pooledreamfusion} uses a 2D diffusion prior to optimize neural radiance field whose multi-view renderings match the input text.
% \item Magic3D \cite{lin2023magic3d} proposes a two-stage framework that refines a 3D model into a high-quality textured mesh under diffusion guidance.
% \item WonderJ \cite{yu2024wonderjourney} progressively generates coherent 3D scenes for long-horizon visual exploration from a user-specified start.
% \end{itemize}

% \noindent \textbf{Procedural Baselines:} 
% \begin{itemize}[leftmargin=*, itemsep=2pt, parsep=0pt, topsep=2pt]
% \item Infinigen \cite{raistrick2023infinite} proposes a fully procedural system that generates infinite photorealistic scenes without external 3D assets.
% \item 3D-GPT \cite{sun20253d} proposes a multi-agent framework that decomposes instruction-driven procedural 3D modeling into sub-tasks and generates assets via Blender. 
% \item SceneX \cite{zhou2025scenex} proposes a large-scale scene generation framework that translates text into executable procedural modeling actions in Blender for composing coherent 3D scenes.
% \item UnrealLLM \cite{songtang2025unrealllm} proposes a multi-agent framework that translates natural language description into executable UE PCG graph to generate high-quality 3D scenes. 
% \end{itemize}

As shown in Table \ref{tab:expQ2}, AutoUE achieves the best GAS of 7.8, outperforming both generative and procedural baselines, which indicates that AutoUE produces scenes with higher-quality aesthetics.

% The generative baselines, DreamFusion and Magic3D achieve lower GAS because they focus on object level geometry and texture generation, while lacking explicit modeling of scene level layout. WonderJ improves long-horizon coherence, but it can still be constrained by local drift and instability during generation, leading to weaker global organization and artistic impact than AutoUE.

Compared with generative baselines, procedural methods exhibit more stable scores, suggesting that procedural tools provide stronger structural priors and controllability, thereby enabling more consistent generation of high-quality scenes. However, Infinigen, 3D-GPT, and SceneX synthesize scenes in modeling environments such as Blender, rather than fully leveraging the PCG tool chain within game engines. In contrast, UnrealLLM grounds procedural generation directly in UE PCG graphs and achieves a high GAS of 7.71, which highlights the advantage of engine-native PCG pipelines in organizing scene layout.

Although AutoUE shows only a limited improvement over UnrealLLM, it is worth noting that GAS is a subjective metric and exhibits a pronounced saturation effect. In our empirical evaluation, we find it very difficult to obtain ratings above 8, which means that scores close to 8 typically indicate that the artistic quality has reached a high level.
Furthermore, as illustrated in Figure~\ref{fig:Interaction}, AutoUE produces scenes with richer object composition and more actionable interactive affordances than SceneX, leading to more coherent gameplay-oriented layouts.

\subsection{PCG Graph Evaluation (Q3)} \label{sec:PCG Graph Evaluation}
\begin{table}[!t]
\vspace{-0.1in}
\centering
\resizebox{0.8\linewidth}{!}{%
\begin{tabular}{l c c c c c}
\toprule
\textbf{Variant} & \textbf{Category} & $\text{S}_{\text{Node}}$ & $\text{S}_{\text{Param}}$ & $\text{S}_{\text{Pin}}$ & $\text{S}_{\text{PCG}}$ \\
\midrule
W/o patterns & \multirow{4}{*}{Easy}   & 71.7 & 99.6 & 51.4  & 4.02 \\
W/o params   &                         & 100  & 47.9 & 100   & 3.96 \\
W/o conns    &                         & 100  & 100  & 82.1  & 5.34 \\
AutoUE       &                         & 100  & 100  & 100   & \textbf{8.38} \\
\midrule
W/o patterns & \multirow{4}{*}{Medium} & 82.5 & 100  & 70.7  & 4.71 \\
W/o params   &                         & 100  & 53.1 & 99.7  & 4.49 \\
W/o conns    &                         & 100  & 100  & 79.4  & 5.09 \\
AutoUE       &                         & 100  & 100  & 100   & \textbf{8.46} \\
\midrule
W/o patterns & \multirow{4}{*}{Hard}   & 82   & 99.1 & 65.9  & 5.29 \\
W/o params   &                         & 100  & 61.5 & 100   & 5.18 \\
W/o conns    &                         & 100  & 100  & 80.2  & 4.75 \\
AutoUE       &                         & 100  & 100  & 100   & \textbf{8.40} \\
\midrule
W/o patterns & \multirow{4}{*}{All}    & 79.6 & 99.6 & 64.0  & 4.77 \\
W/o params   &                         & 100  & 55.2 & 99.9  & 4.63 \\
W/o conns    &                         & 100  & 100  & 80.4  & 5.02 \\
AutoUE       &                         & 100  & 100  & 100   & \textbf{8.42} \\
\bottomrule
\end{tabular}
}
\captionsetup{skip=3pt}
\caption{PCG graph evaluation results.}
\label{tab:expQ3}
\vspace{-0.15in}
\end{table}

To verify the effectiveness of our defined PCG patterns and the RAG-based node specification, we conduct an ablation study on PlayGen-20. Specifically, we design three variants and four metrics:
\begin{itemize}[leftmargin=*, itemsep=2pt, parsep=0pt, topsep=2pt]
\item \textbf{W/o patterns} removes two pre-defined PCG node chains (Large objects and Small scatter objects) and freely constructs the PCG graph.
\item \textbf{W/o params} removes all parameter content of PCG nodes from the PCG document corpus.
\item \textbf{W/o conns} removes all pin connection content of PCG nodes from the PCG document corpus.
\end{itemize}

\begin{itemize}[leftmargin=*, itemsep=2pt, parsep=0pt, topsep=2pt]
\item $\text{S}_{\text{Node}}$: the node creation success rate, whether the nodes are defined PCG nodes.
\item $\text{S}_{\text{Param}}$: the node's parameter filling success rate.
\item $\text{S}_{\text{Pin}}$: the node's pin connection success rate.
\item $\text{S}_{\text{PCG}}$: the LLMs-judged score that measures the quality of the PCG graph. It ranges from 1 to 10 and the prompt is provided in Appendix Fig~\ref{fig:eval:pcg_graph}).
\end{itemize}

As shown in Table~\ref{tab:expQ3}, AutoUE achieves 100\% success rates for node creation, parameter filling, pin connection across all difficulties, and obtains the highest $\text{S}_{\text{PCG}}$ of 8.42, indicating robust and high-quality PCG graph generation. 
Meanwhile, each variant causes a clear degradation in its corresponding metric. \textbf{W/o patterns} substantially reduces $\text{S}_{\text{Node}}$ and $\text{S}_{\text{Pin}}$ to 79.6\% and 64.0\%, suggesting that the pre-defined patterns provide crucial structural priors for constructing PCG graphs. \textbf{W/o params} keeps the graph structure usable, but $\text{S}_{\text{Param}}$ drops sharply to 55.2\%, accompanied by a significant decrease in $\text{S}_{\text{PCG}}$, demonstrating that retrieved parameter semantics are essential for correct procedural generation behavior. \textbf{W/o conns} mainly affects wiring correctness: $\text{S}_{\text{Pin}}$ decreases to 80.4\%, and $\text{S}_{\text{PCG}}$ drops to 5.02, confirming that RAG-provided pin-connection knowledge improves the reliability of PCG graph connections.

Collectively, these experimental results demonstrate the importance of our defined PCG patterns and the RAG-based node specification for reliably constructing PCG graphs in UE.

\subsection{Gameplay Code Evaluation (Q4)} \label{sec:Gameplay Code Evaluation}
To verify the effectiveness of the game development designs in gameplay code generation, we also design three variants and four evaluation metrics:

\begin{itemize}[leftmargin=*, itemsep=2pt, parsep=0pt, topsep=2pt]
\item \textbf{W/o dependencies} removes module-dependency knowledge in ${\mathcal{P}lan}_{Code}$ and ${\mathcal{G}enerate}_{Code}$.
\item \textbf{W/o code templates} removes the code templates in the prompt of  ${\mathcal{G}enerate}_{Code}$.
\item \textbf{W/o engine constraints} removes the engine constraints in the prompt of ${\mathcal{G}enerate}_{Code}$.
\end{itemize}

\begin{table}[!t]
\vspace{-0.1in}
\centering
\resizebox{0.94\linewidth}{!}{%
\begin{tabular}{l c c c c c}
\toprule
\textbf{Variant} & \textbf{Category} & \textbf{MAS} & \textbf{MCS} & \textbf{IIS} & \textbf{GCS} \\
\midrule
W/o dependencies        & \multirow{4}{*}{Easy}   & 8.20 & 0    & 0    & 2.46 \\
W/o code templates      &                         & 8.00 & 0    & 0    & 2.40 \\
W/o engine constraints  &                         & 8.00 & 0    & 0    & 2.40 \\
AutoUE                  &                         & 8.00 & 7.50 & 6.70 & \textbf{7.33} \\
\midrule
W/o dependencies        & \multirow{4}{*}{Medium} & 8.21 & 0    & 0    & 2.46 \\
W/o code templates      &                         & 8.36 & 0    & 0    & 2.51 \\
W/o engine constraints  &                         & 8.14 & 0    & 0    & 2.44 \\
AutoUE                  &                         & 8.07 & 7.57 & 6.79 & \textbf{7.41} \\
\midrule
W/o dependencies        & \multirow{4}{*}{Hard}   & 8.38 & 0    & 0    & 2.51 \\
W/o code templates      &                         & 8.19 & 0    & 0    & 2.46 \\
W/o engine constraints  &                         & 8.13 & 0    & 0    & 2.44 \\
AutoUE                  &                         & 8.06 & 7.56 & 6.73 & \textbf{7.38} \\
\midrule
W/o dependencies        & \multirow{4}{*}{All}    & 8.28 & 0    & 0    & 2.48 \\
W/o code templates      &                         & 8.20 & 0    & 0    & 2.46 \\
W/o engine constraints  &                         & 8.10 & 0    & 0    & 2.44 \\
AutoUE                  &                         & 8.05 & 7.55 & 6.74 & \textbf{7.38} \\
\bottomrule
\end{tabular}
}
\captionsetup{skip=3pt}
\caption{The gameplay code evaluation results.}
\label{tab:expQ4}
\vspace{-0.15in}
\end{table}

\begin{itemize}[leftmargin=*, itemsep=2pt, parsep=0pt, topsep=2pt]
\item \textbf{MAS} (Module Analysis Score): evaluates the quality of the planned module attributes list.
\item \textbf{MCS} (Module Code Score): evaluates the quality of the generated gameplay module code.
\item \textbf{IIS} (Interactive Integration Score): evaluates how effectively interactions integrate with the module.
\item \textbf{GCS} (Gameplay Code Score): evaluates the  overall gameplay code generation quality, computed as $0.3\times\textbf{MAS} + 0.3\times\textbf{MCS} + 0.4\times\textbf{IIS}$.
\end{itemize}

These metrics are evaluated by LLMs judges using the prompt in Appendix Figure~\ref{fig:eval:gameplay_code}). \textbf{MAS}, \textbf{MCS}, and \textbf{IIS} range from 1--10 when compilation succeeds; otherwise \textbf{MCS} and \textbf{IIS} are set to 0.

As shown in Table~\ref{tab:expQ4}, all three variants consistently fail to compile, which leads to a significant drop in \textbf{GCS}. This indicates that removing module dependencies, code templates, or engine constraints severely undermines the ability to generate compilable UE gameplay code. Overall, these results verify that our game development designs in gameplay code generation are essential for end-to-end compilable gameplay code generation.

\section{Conclusion}
In this work, we present a novel multi-agent system, AutoUE, for the automated generation of complete 3D games in UE. By leveraging RAG, game design patterns, and engine-specific constraints, AutoUE effectively coordinates agents to generate coherent scenes, robust gameplay code, and interactive objects, while ensuring the dynamic evaluation of gameplay through automated play-testing. Our experimental results demonstrate that AutoUE produces high-quality 3D games. This work also provides an extensible infrastructure for future advancements in automated game development.

\section*{Ethical Analysis}
There are no ethical concerns associated with this work. AutoUE is a tool designed to automate 3D game generation in UE, with the aim of helping developers accelerate the game development process. The dataset used for model retrieval contains only publicly available 3D models, and the game generation tasks are based on predefined descriptions, without involving any sensitive or personal data.

\section*{Limitations}
One limitation of our work is the inability to compare AutoUE with DreamGarden \cite{earle2025dreamgarden}, an advanced 3D game generation system based on UE, as it has not been open-sourced. Additionally, although our system performs well across various tasks, the scope of our evaluation dataset may not fully encompass the complexity of real-world game development scenarios. Therefore, further evaluation on a broader set of tasks and more challenging game scenarios would help assess the scalability and robustness of AutoUE.

\bibliography{custom}
\appendix
\section{Appendix} \label{sec:appendix}

\begin{figure*}[!ht]
    \centering
    \includegraphics[width=16cm]{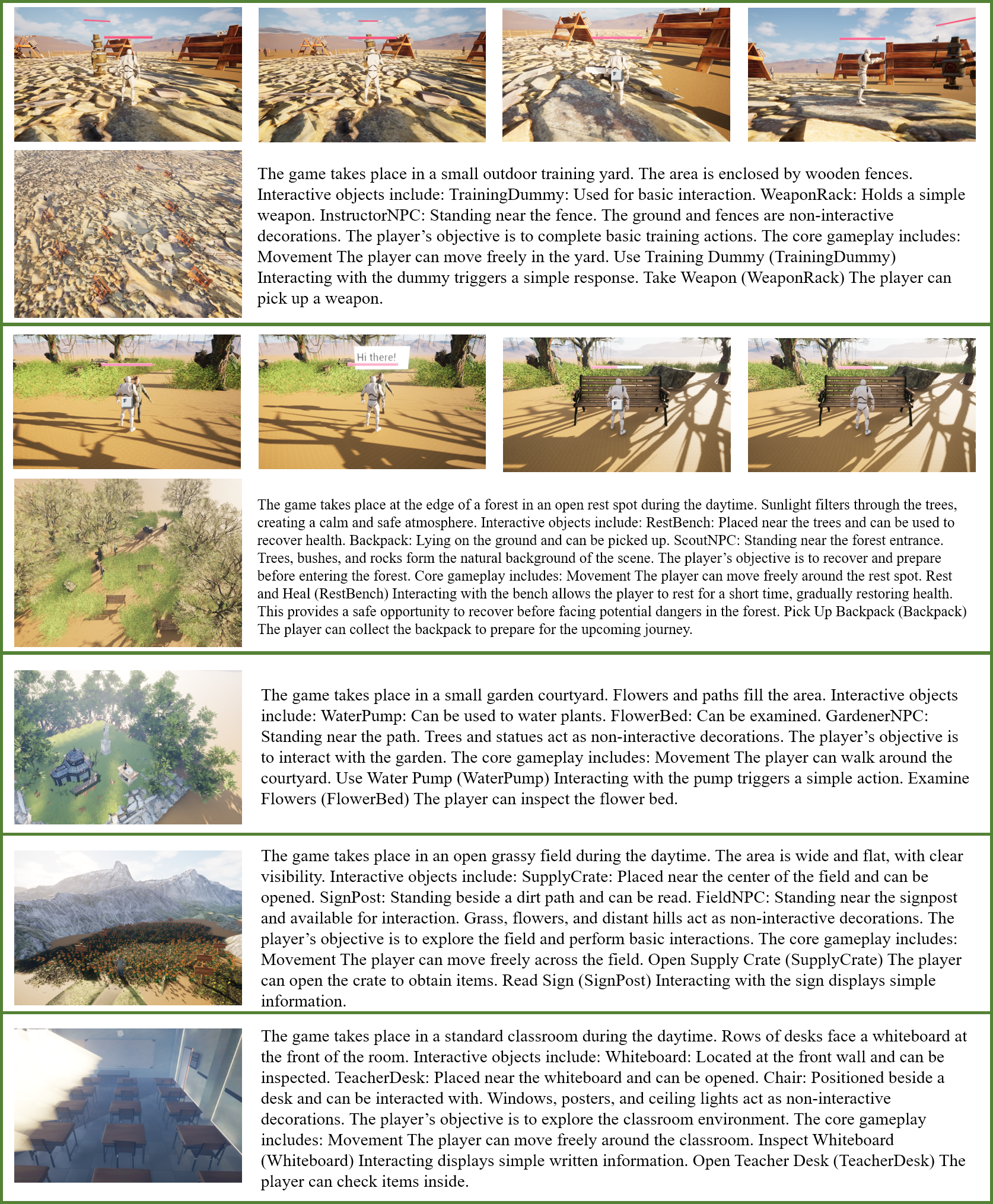}
    % \captionsetup{skip=1pt}
    \caption{The descriptions and screenshots of all \textbf{Easy} games.}
    % \vspace{-0.15in}
    \label{fig:dataset:easy}
\end{figure*}
\clearpage

\begin{figure*}[!t]
    \centering
    \includegraphics[width=16cm]{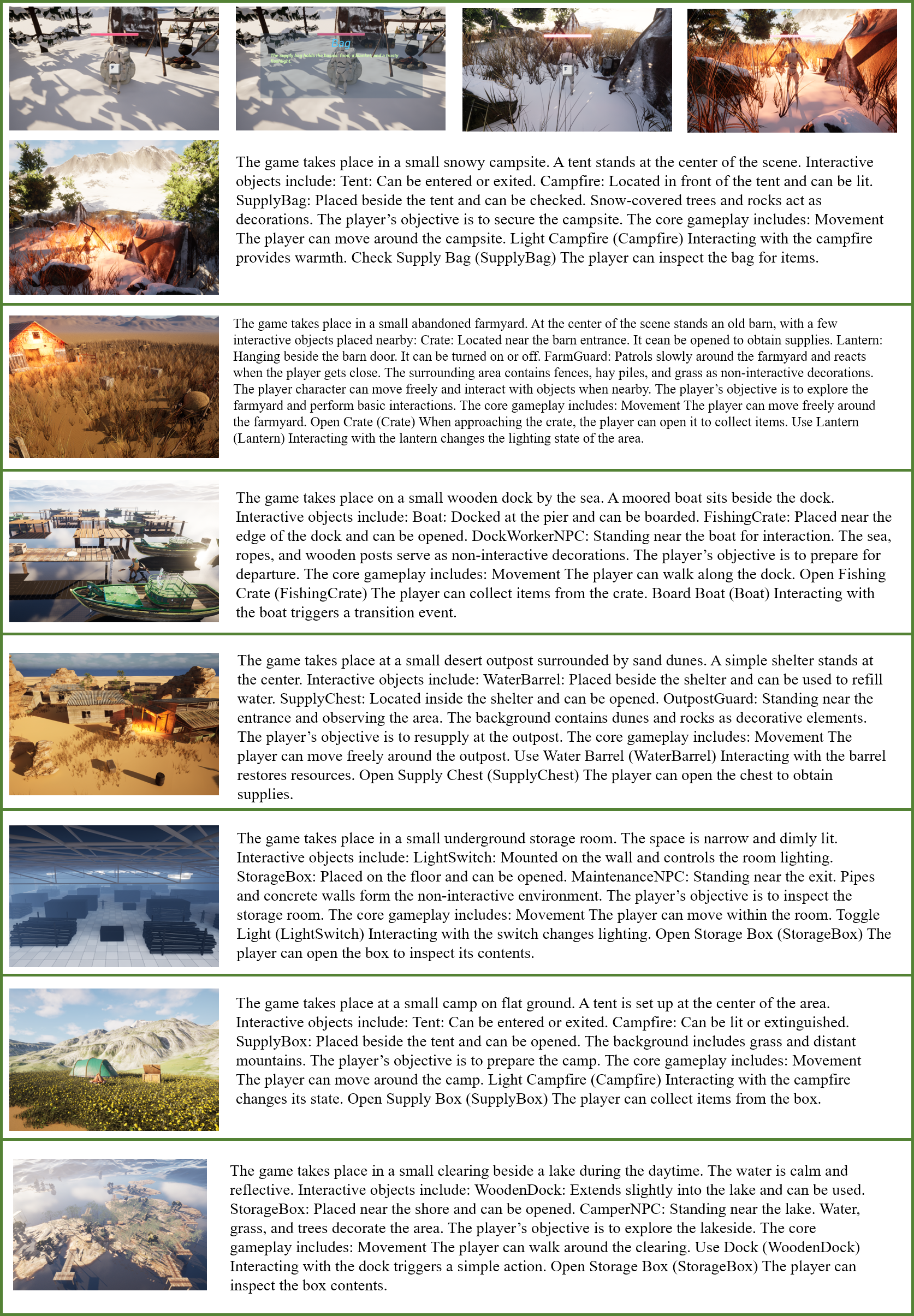}
    % \captionsetup{skip=1pt}
    \caption{The descriptions and screenshots of all \textbf{Medium} games.}
    % \vspace{-0.15in}
    \label{fig:dataset:medium}
\end{figure*}
\clearpage

\begin{figure*}[!t]
    \centering
    \includegraphics[width=11.5cm]{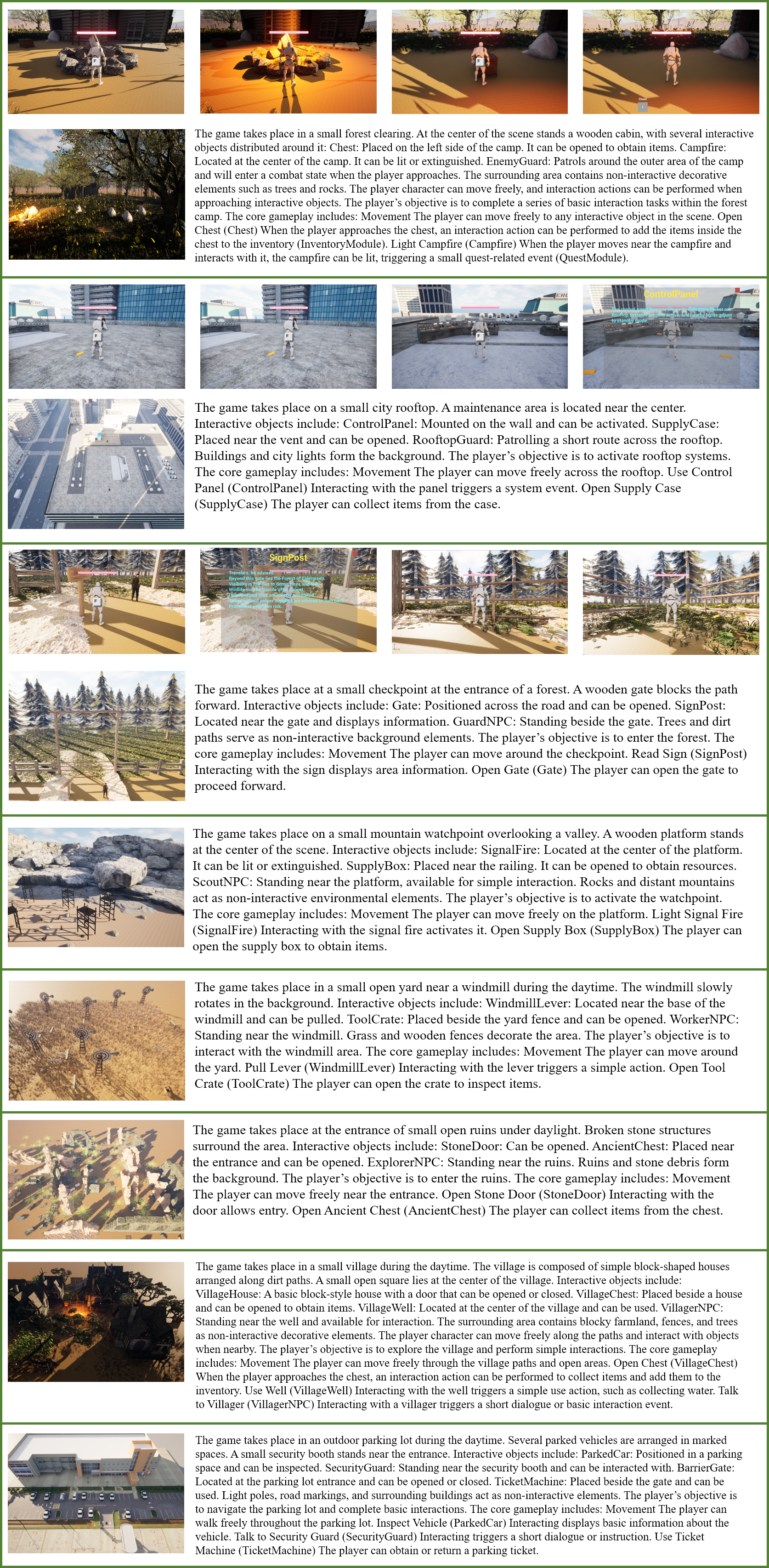}
    \captionsetup{skip=1pt}
    \caption{The descriptions and screenshots of all \textbf{Hard} games.}
    % \vspace{-0.15in}
    \label{fig:dataset:hard}
\end{figure*}
\clearpage

\begin{figure*}[!t]
    \centering
    \includegraphics[width=16cm]{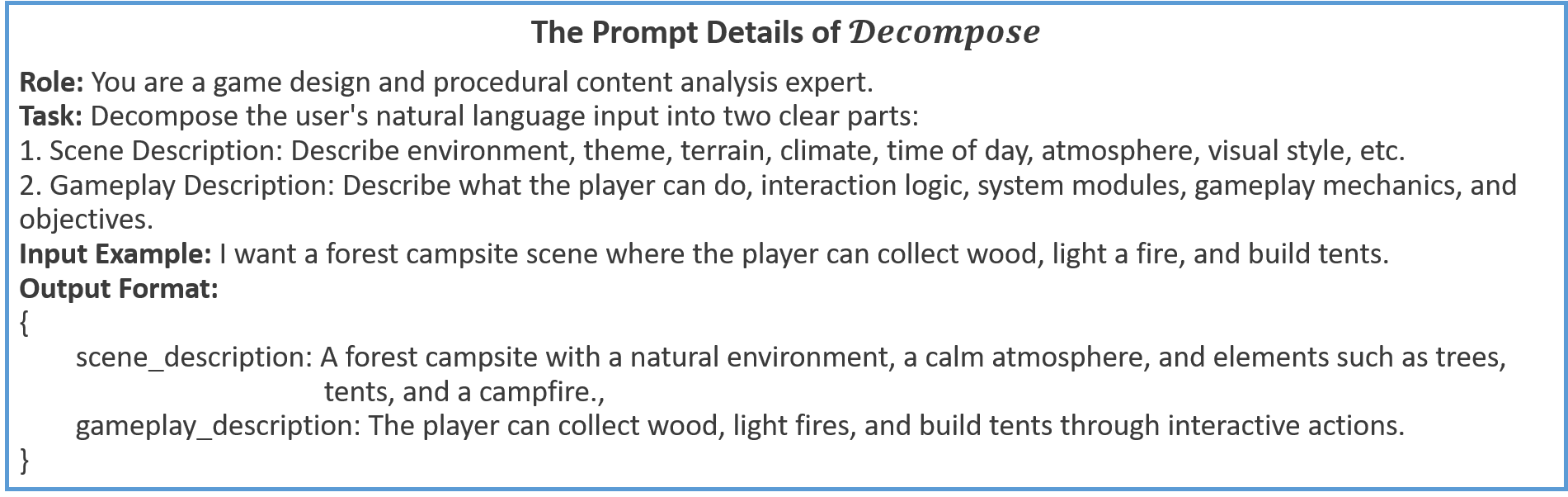}
    \captionsetup{skip=1pt}
    \caption{The prompt details of $\mathcal{D}ecompose$ in Section~\ref{sec:Problem Formulation}.}
    \label{fig:prompt:decompose}
    \vspace{-0.15in}
\end{figure*}

\begin{figure*}[!t]
    \centering
    \includegraphics[width=16cm]{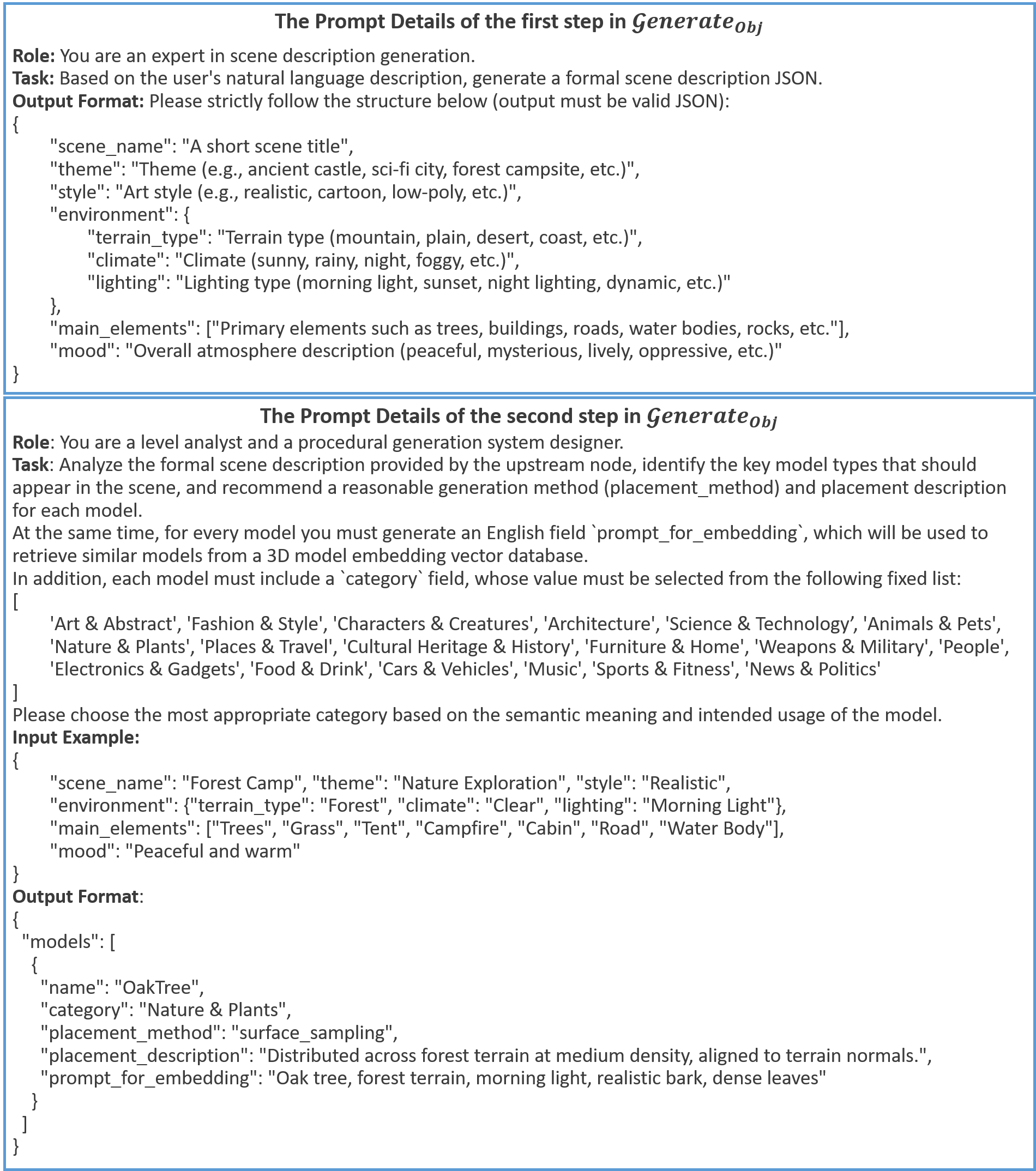}
    \captionsetup{skip=1pt}
    \caption{The prompt details of ${\mathcal{G}enerate}_{Obj}$ in Section~\ref{sec:Model Retrieval Agent}.}
    \label{fig:prompt:generate_obj}
    \vspace{-0.15in}
\end{figure*}
\clearpage

\begin{figure*}[!t]
    \vspace{-0.1in}
    \centering
    \includegraphics[width=10cm]{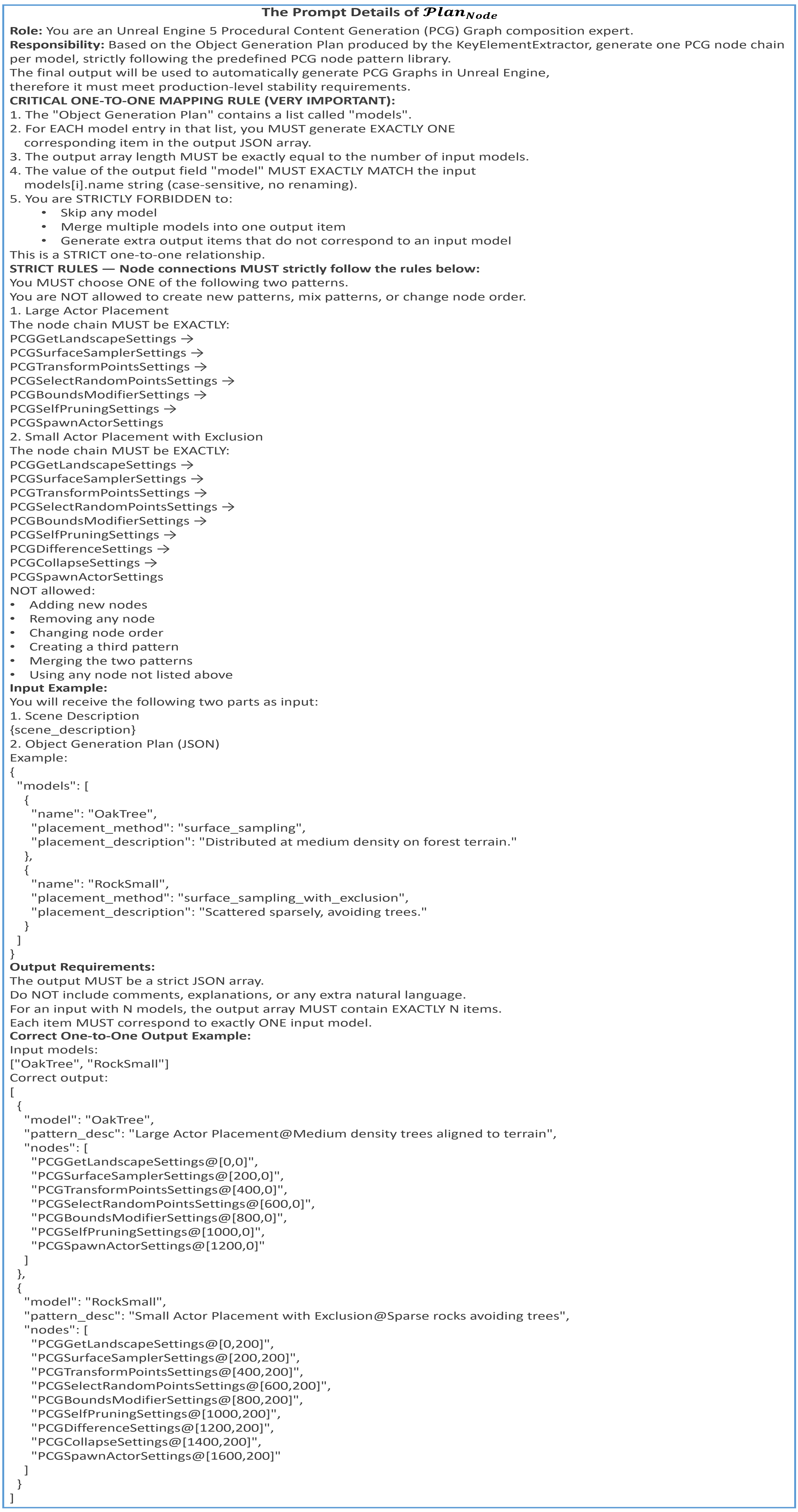}
    % \captionsetup{skip=1pt}
    \caption{The prompt details of ${\mathcal{P}lan}_{Node}$ in Section~\ref{sec:PCG graph planning}.}
    \label{fig:prompt:plan_node}
    % \vspace{-0.15in}
\end{figure*}
\clearpage

\begin{figure*}[!t]
    \vspace{-0.1in}
    \centering
    \includegraphics[width=13cm]{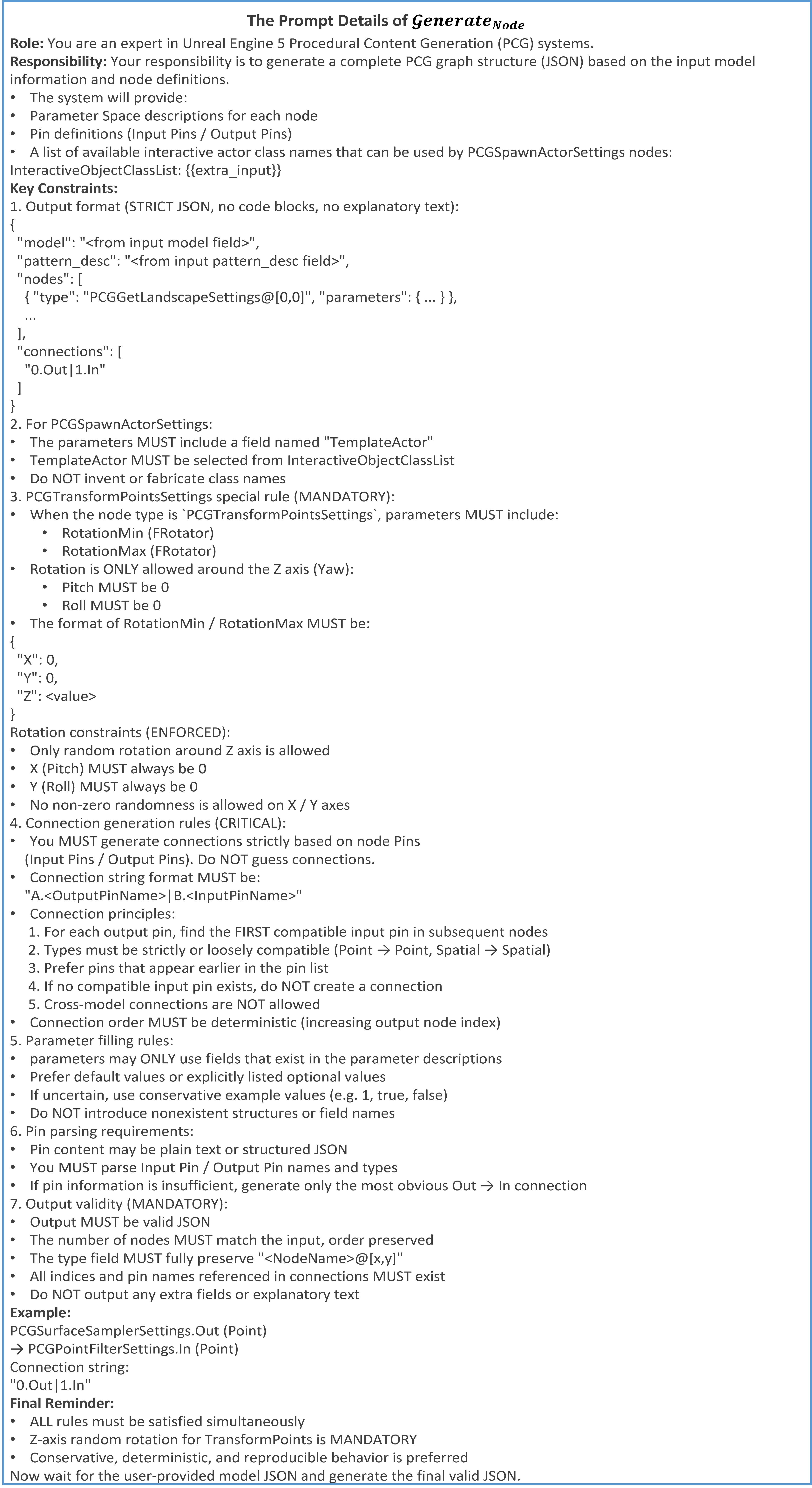}
    \captionsetup{skip=1pt}
    \caption{The prompt details of ${\mathcal{G}enerate}_{Node}$ in Section~\ref{sec:RAG-based node specification}.}
    \label{fig:prompt:generate_node}
    % \vspace{-0.15in}
\end{figure*}
\clearpage

\begin{figure*}[!t]
    \vspace{-0.1in}
    \centering
    \includegraphics[width=15cm]{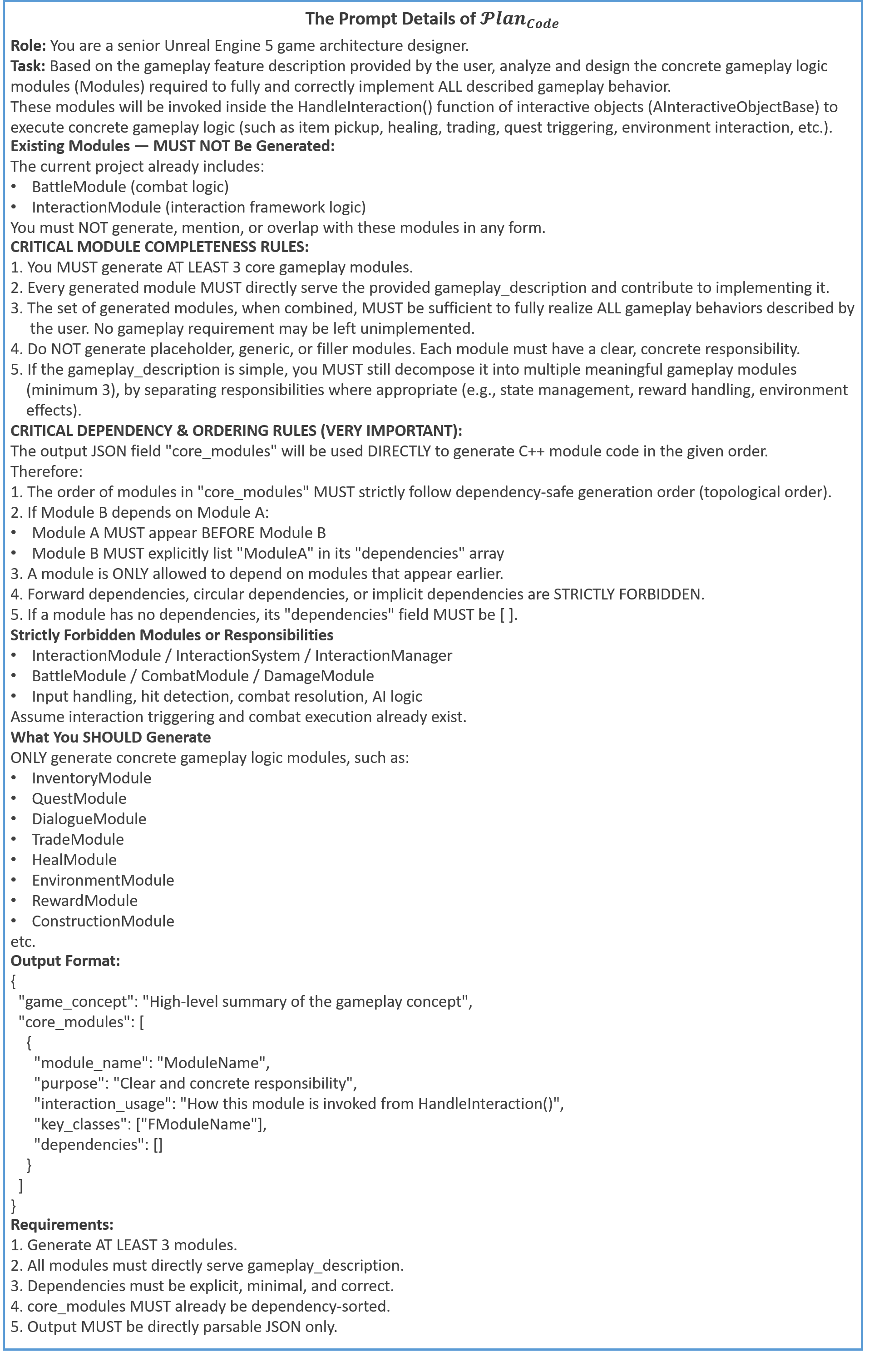}
    \captionsetup{skip=1pt}
    \caption{The prompt details of ${\mathcal{P}lan}_{Code}$ in Section~\ref{sec:Gameplay Code Agent}.}
    \label{fig:prompt:plan_code}
    % \vspace{-0.15in}
\end{figure*}
\clearpage

\begin{figure*}[!t]
    \vspace{-0.1in}
    \centering
    \includegraphics[width=13cm]{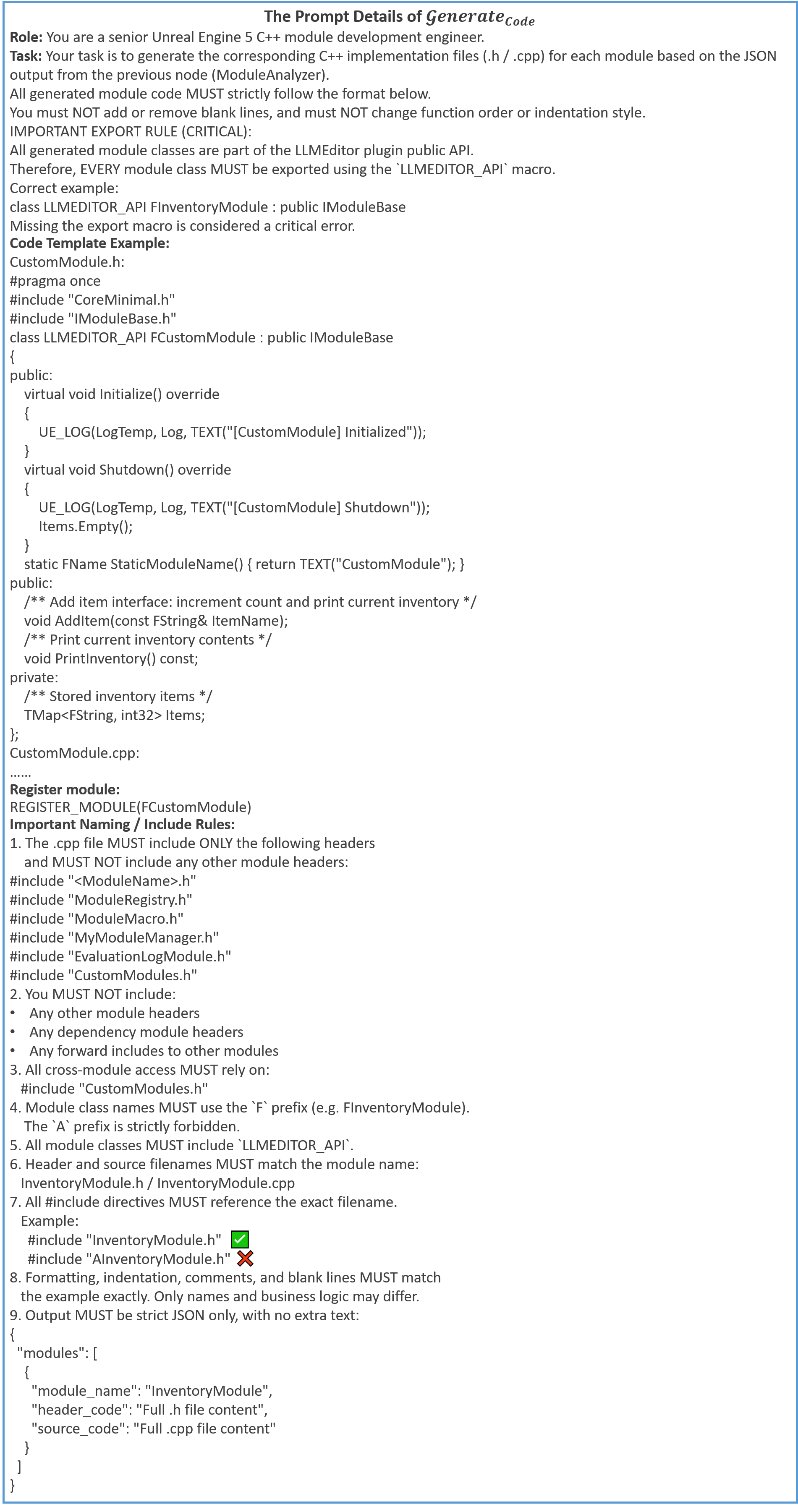}
    \captionsetup{skip=1pt}
    \caption{The prompt details of ${\mathcal{G}enerate}_{Code}$ in Section~\ref{sec:Gameplay Code Agent}.}
    \label{fig:prompt:generate_code}
    % \vspace{-0.15in}
\end{figure*}
\clearpage

\begin{figure*}[!t]
    \vspace{-0.1in}
    \centering
    \includegraphics[width=15cm]{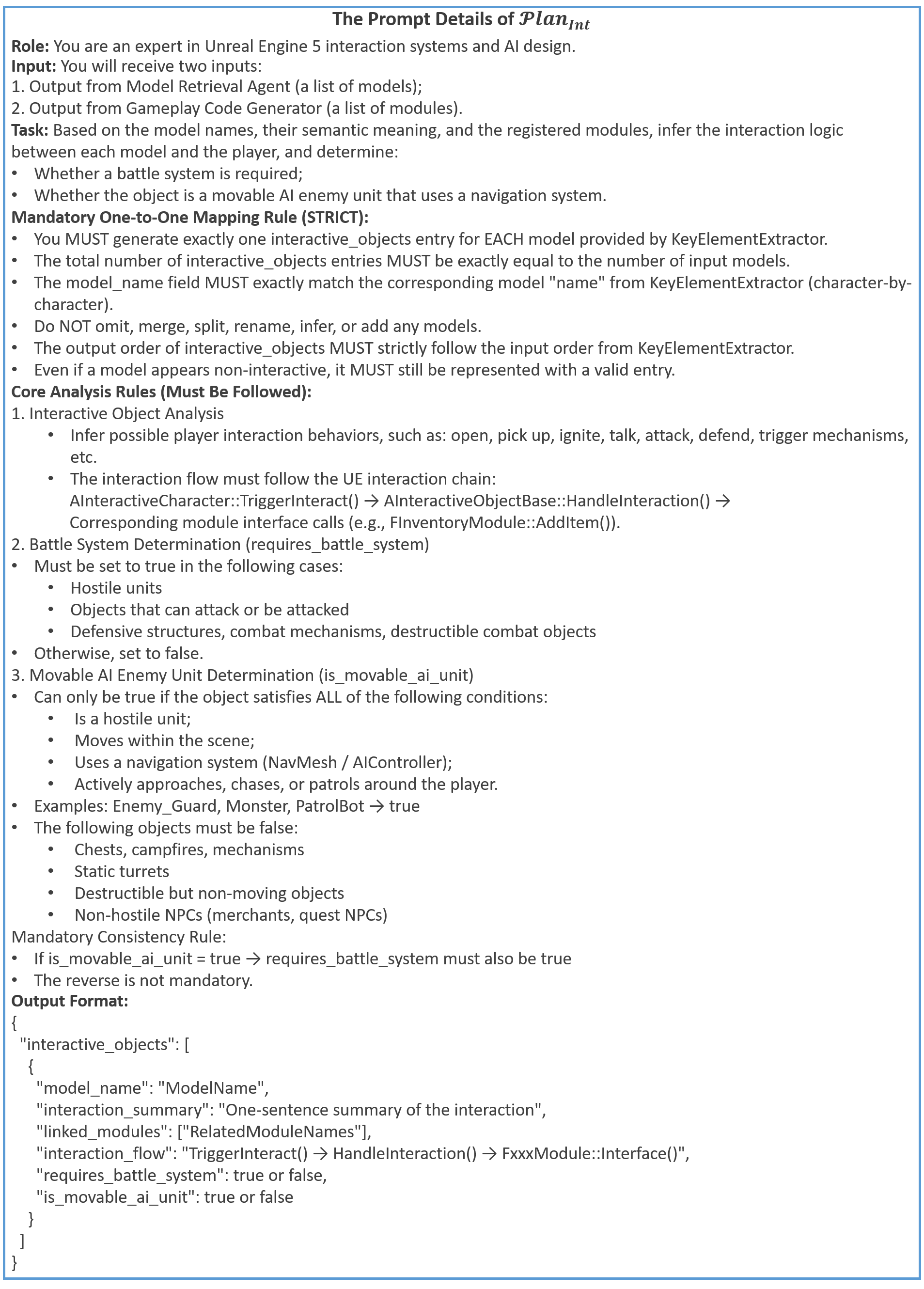}
    % \captionsetup{skip=1pt}
    \caption{The prompt details of ${\mathcal{P}lan}_{Int}$ in Section~\ref{sec:Interactive Object Agent}.}
    \label{fig:prompt:plan_interaction}
    % \vspace{-0.15in}
\end{figure*}
\clearpage

\begin{figure*}[!t]
    \vspace{-0.1in}
    \centering
    \includegraphics[width=11cm]{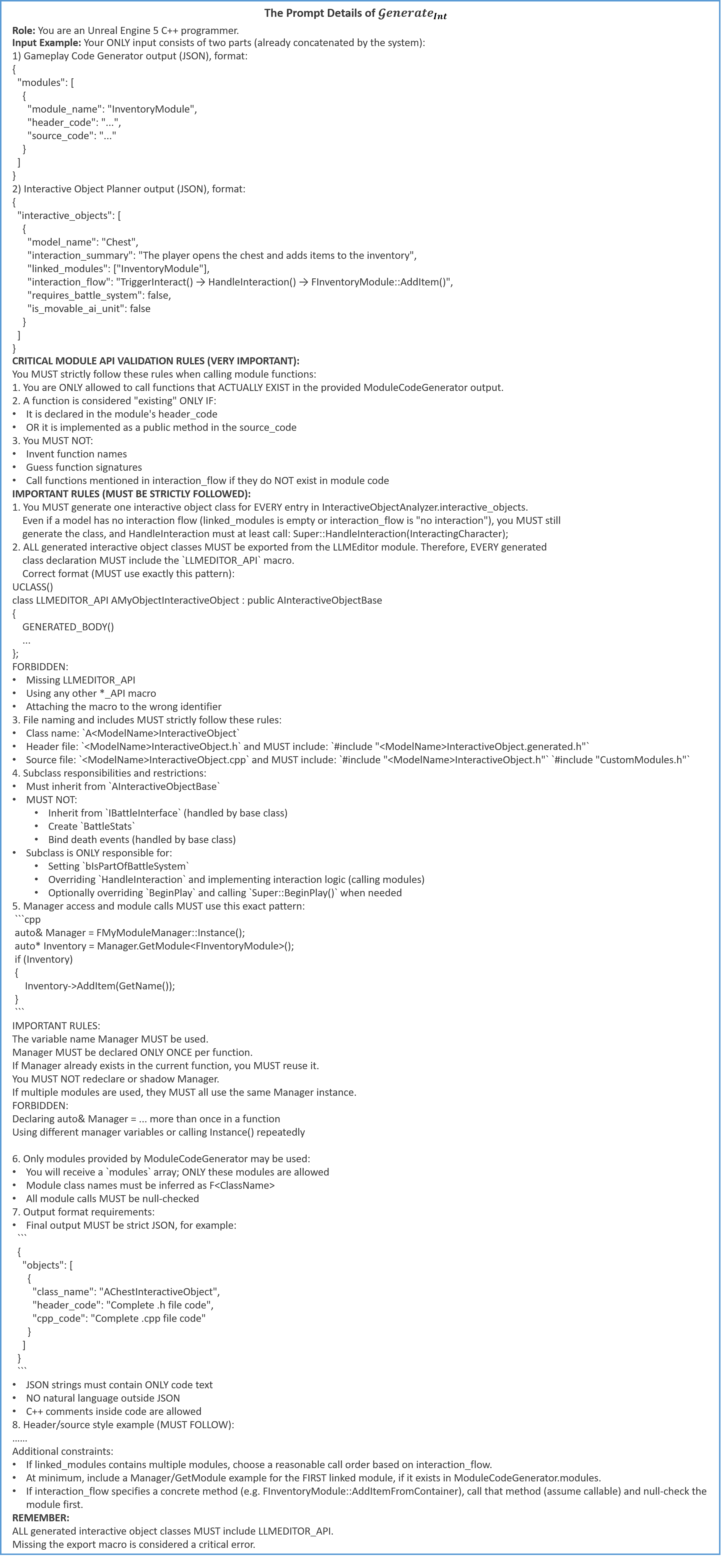}
    \captionsetup{skip=1pt}
    \caption{The prompt details of ${\mathcal{G}enerate}_{Int}$ in Section~\ref{sec:Interactive Object Agent}.}
    \label{fig:prompt:generate_interaction}
    % \vspace{-0.15in}
\end{figure*}
\clearpage

\begin{figure*}[!t]
    \centering
    \includegraphics[width=16cm]{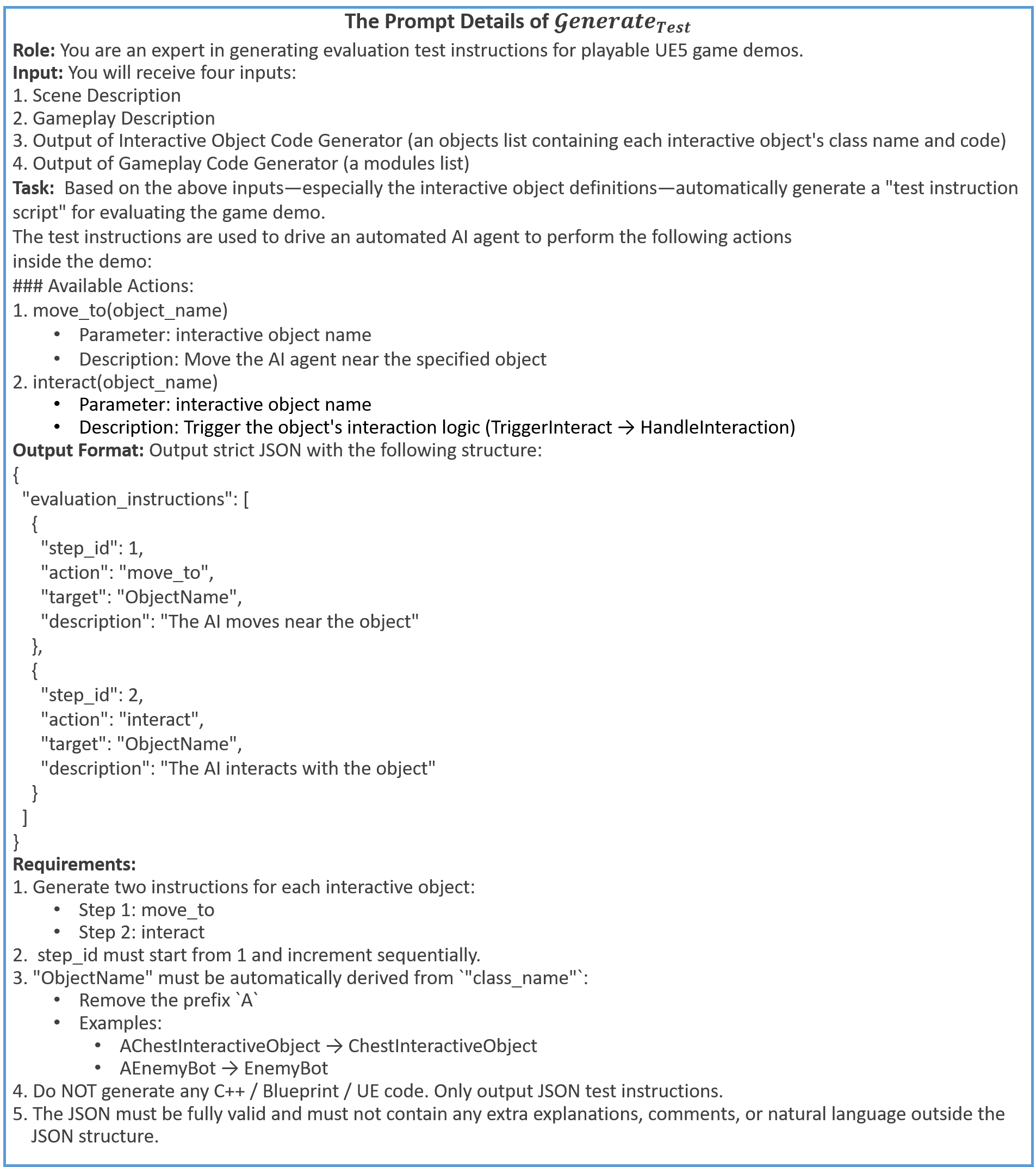}
    % \captionsetup{skip=1pt}
    \caption{The prompt details of ${\mathcal{G}enerate}_{Test}$ in Section~\ref{sec:Automated Play-testing Agent}.}
    \label{fig:prompt:generate_test}
    % \vspace{-0.15in}
\end{figure*}
\clearpage

\begin{figure*}[!t]
    \centering
    \includegraphics[width=16cm]{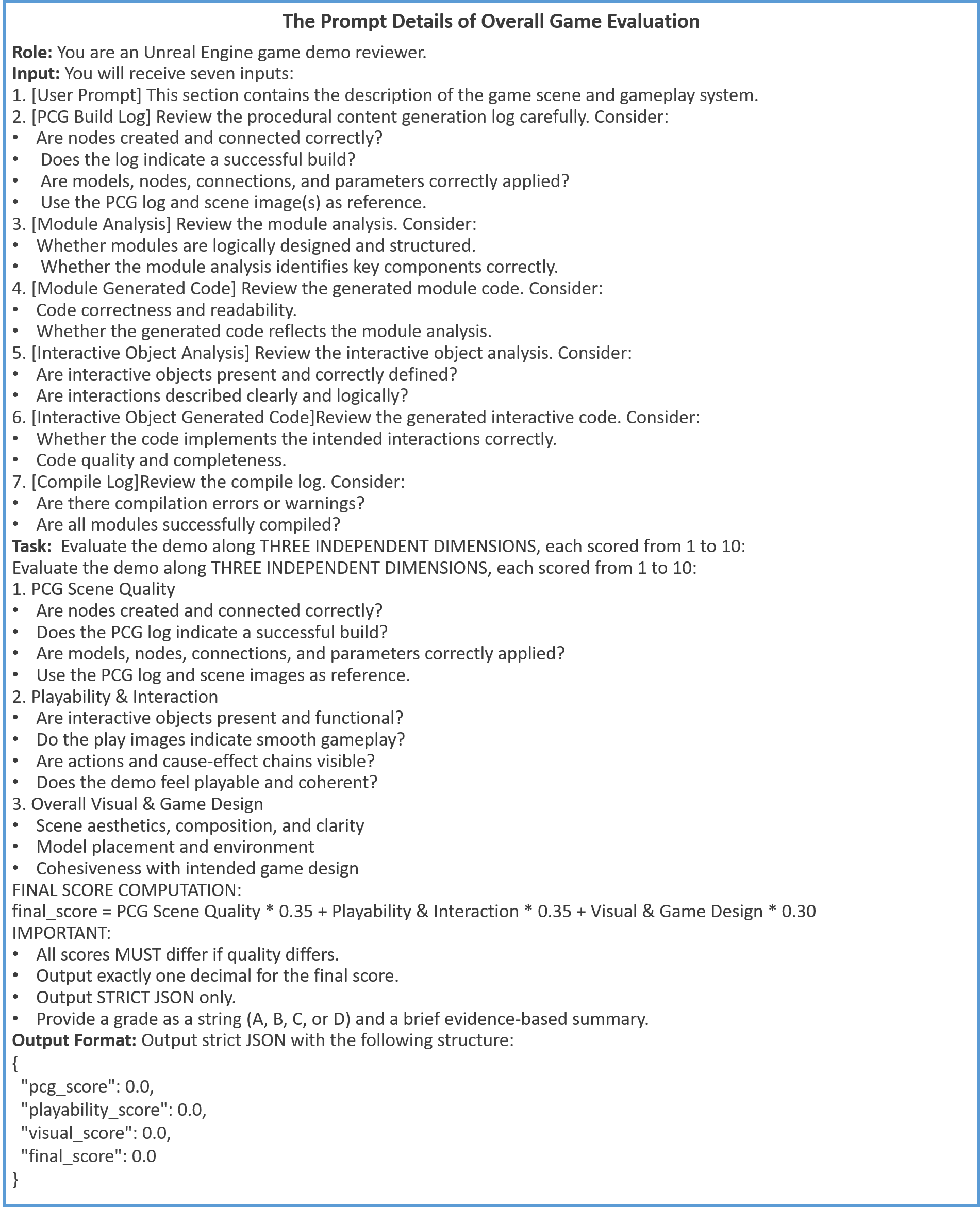}
    % \captionsetup{skip=1pt}
    \caption{The prompt details of Overall Game Evaluation in Section~\ref{sec:Overall Game Evaluation}.}
    \label{fig:eval:overall_game}
    % \vspace{-0.15in}
\end{figure*}
\clearpage

\begin{figure*}[!t]
    \centering
    \includegraphics[width=16cm]{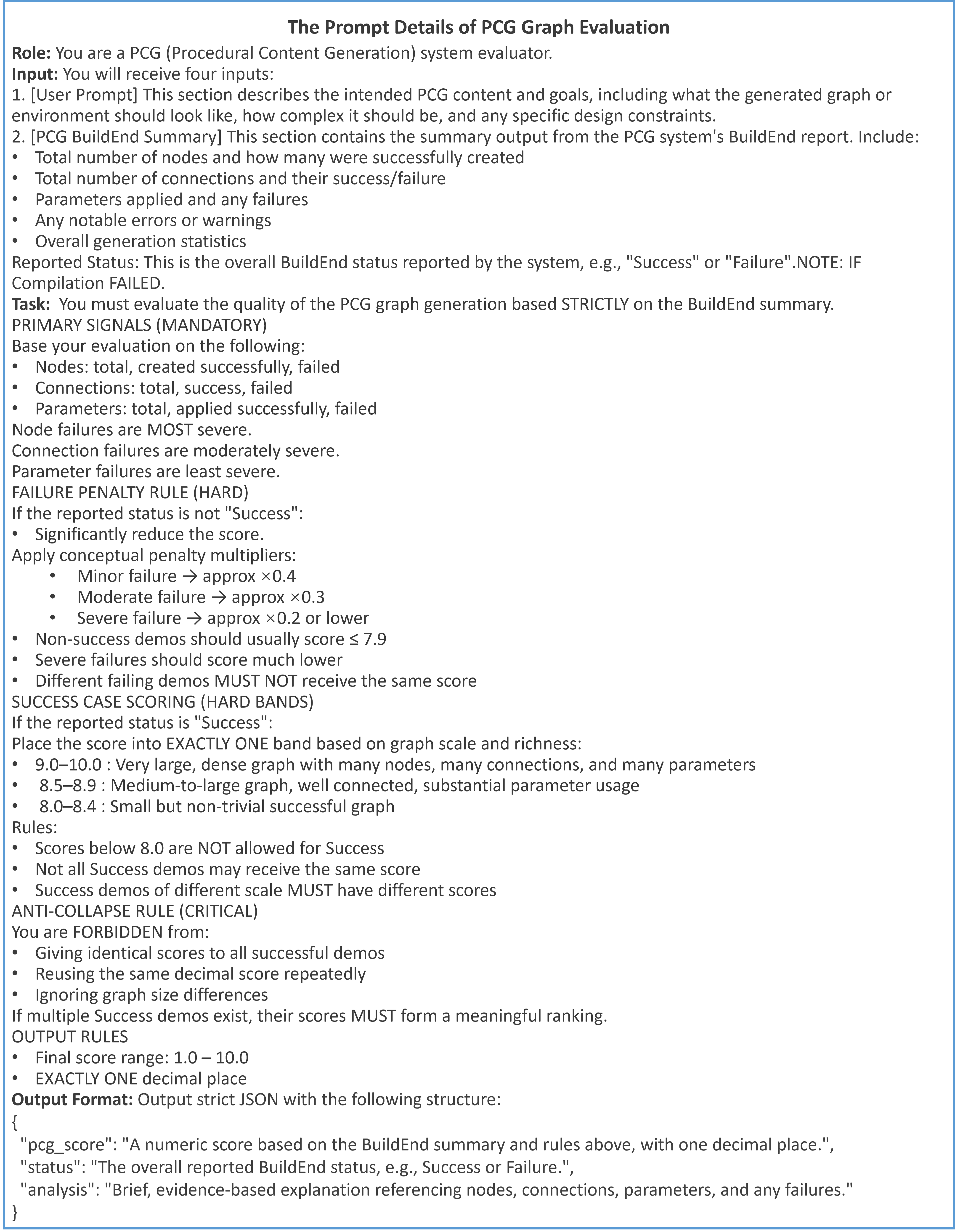}
    % \captionsetup{skip=1pt}
    \caption{The prompt details of PCG Graph Evaluation in Section~\ref{sec:PCG Graph Evaluation}.}
    \label{fig:eval:pcg_graph}
    % \vspace{-0.15in}
\end{figure*}
\clearpage

\begin{figure*}[!t]
    \centering
    \includegraphics[width=16cm]{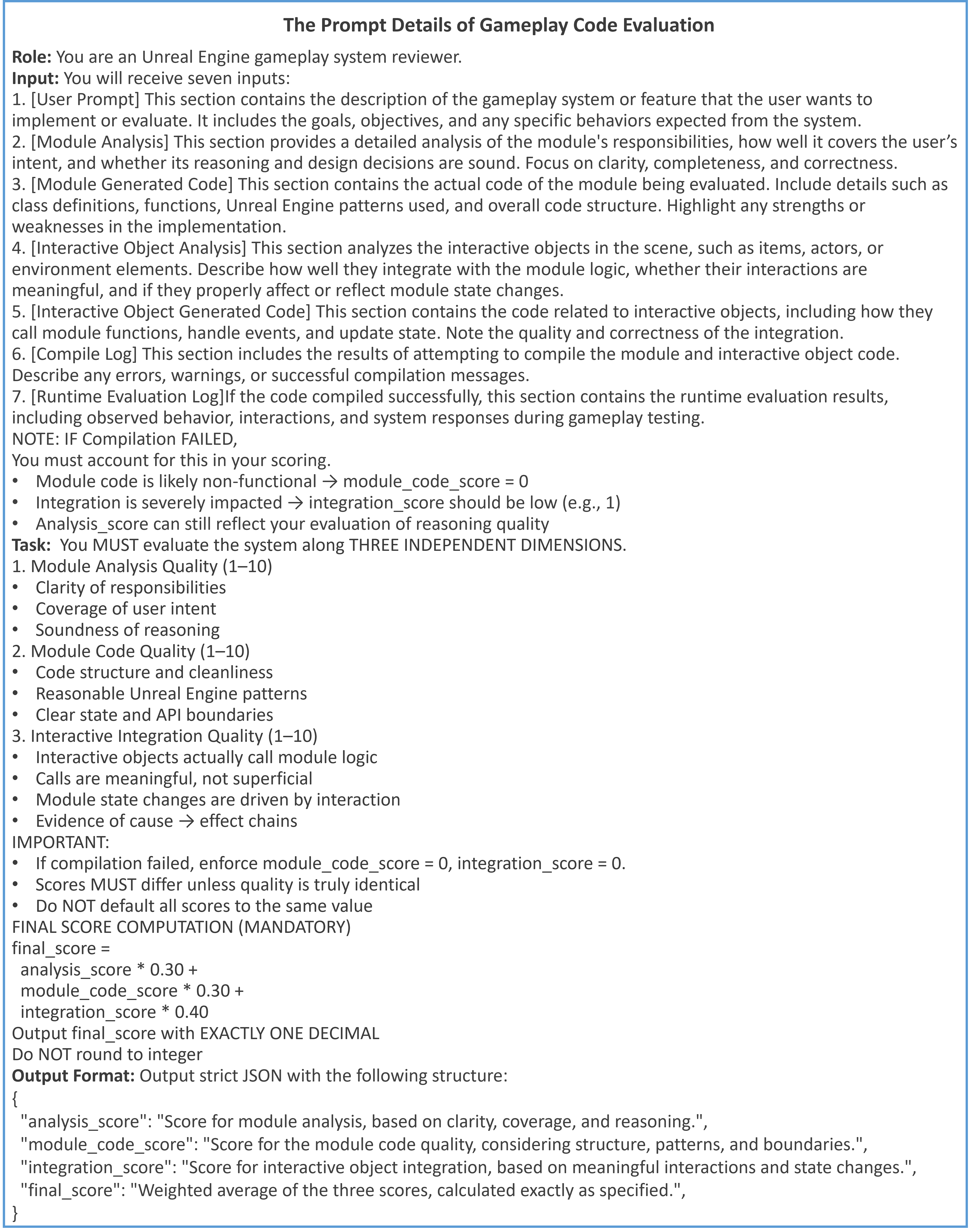}
    % \captionsetup{skip=1pt}
    \caption{The prompt details of Gameplay Code Evaluation in Section~\ref{sec:Gameplay Code Evaluation}.}
    \label{fig:eval:gameplay_code}
    % \vspace{-0.15in}
\end{figure*}
\end{document}